\RequirePackage{fix-cm}
\documentclass[smallextended]{svjour3}
\makeatletter\if@twocolumn\PassOptionsToPackage{switch}{lineno}\else\fi\makeatother

\usepackage{amsfonts,amssymb,amsbsy,tabulary,amsmath}
\smartqed  
\usepackage{graphicx}
\usepackage{endfloat}

\usepackage{url,multirow,morefloats,floatflt,cancel,tfrupee}
\makeatletter

\AtBeginDocument{\@ifpackageloaded{textcomp}{}{\usepackage{textcomp}}}
\makeatother
\usepackage{colortbl}
\usepackage{xcolor}
\usepackage{pifont}
\usepackage[nointegrals]{wasysym}
\makeatletter

\def\mcWidth#1{\csname TY@F#1\endcsname+\tabcolsep}

\def\cAlignHack{\rightskip\@flushglue\leftskip\@flushglue\parindent\z@\parfillskip\z@skip}
\def\rAlignHack{\rightskip\z@skip\leftskip\@flushglue \parindent\z@\parfillskip\z@skip}

\@ifundefined{etal}{}{}

\usepackage{ifxetex}
\ifxetex\else\if@twocolumn\@ifpackageloaded{stfloats}{}{\usepackage{dblfloatfix}}\fi\fi

\AtBeginDocument{
\expandafter\ifx\csname eqalign\endcsname\relax
\def\eqalign#1{\null\vcenter{\def\\{\cr}\openup\jot\m@th
  \ialign{\strut$\displaystyle{##}$\hfil&$\displaystyle{{}##}$\hfil
      \crcr#1\crcr}}\,}
\fi
}

\AtBeginDocument{%
  \@ifpackageloaded{endfloat}%
   {\renewcommand\efloat@iwrite[1]{\immediate\expandafter\protected@write\csname efloat@post#1\endcsname{}}}{\newif\ifefloat@tables}%
}%

\def\BreakURLText#1{\@tfor\brk@tempa:=#1\do{\brk@tempa\hskip0pt}}
\let\lt=<
\let\gt=>
\def\processVert{\ifmmode|\else\textbar\fi}

\@ifundefined{subparagraph}{
\def\subparagraph{\@startsection{paragraph}{5}{2\parindent}{0ex plus 0.1ex minus 0.1ex}%
{0ex}{\normalfont\small\itshape}}%
}{}

\newcommand\role[1]{\unskip}
\newcommand\aucollab[1]{\unskip}
  
\@ifundefined{tsGraphicsScaleX}{\gdef\tsGraphicsScaleX{1}}{}
\@ifundefined{tsGraphicsScaleY}{\gdef\tsGraphicsScaleY{.9}}{}
\def\checkGraphicsWidth{\ifdim\Gin@nat@width>\linewidth
	\tsGraphicsScaleX\linewidth\else\Gin@nat@width\fi}

\def\checkGraphicsHeight{\ifdim\Gin@nat@height>.9\textheight
	\tsGraphicsScaleY\textheight\else\Gin@nat@height\fi}

\def\fixFloatSize#1{}
\let\ts@includegraphics\includegraphics

\def\inlinegraphic[#1]#2{{\edef\@tempa{#1}\edef\baseline@shift{\ifx\@tempa\@empty0\else#1\fi}\edef\tempZ{\the\numexpr(\numexpr(\baseline@shift*\f@size/100))}\protect\raisebox{\tempZ pt}{\ts@includegraphics{#2}}}}

\AtBeginDocument{\def\includegraphics{\@ifnextchar[{\ts@includegraphics}{\ts@includegraphics[width=\checkGraphicsWidth,height=\checkGraphicsHeight,keepaspectratio]}}}

\DeclareMathAlphabet{\mathpzc}{OT1}{pzc}{m}{it}

\def\URL#1#2{\@ifundefined{href}{#2}{\href{#1}{#2}}}

\def\UrlOrds{\do\*\do\-\do\~\do\'\do\"\do\-}%
\g@addto@macro{\UrlBreaks}{\UrlOrds}

\usepackage{epsfig}

\edef\fntEncoding{\f@encoding}

\makeatother

\newif\ifmultipleabstract\multipleabstractfalse%
\usepackage[authoryear]{natbib}

\usepackage[T1]{fontenc}

\makeatletter	

\AtBeginDocument{%
  \@ifpackageloaded{endfloat}%
   {
\renewcommand*\efloat@process[2]{%
  \ef@ifct{#1}{%
    \expandafter\immediate\expandafter\closeout\csname efloat@post#1\endcsname
    \ef@setct{#1}{0}%
    \clearpage                                                         
        
    \efloat@ifflag{#2list}{
      {\normalsize\efloat@listof{#2}}
    }{}%

    \efloat@ifflag{#2head}{%
      \section*{\@nameuse{#2section}}
      \suppressfloats[t]
    }{}

    \markboth                                                          
      {\expandafter\uppercase\expandafter{\csname #2section\endcsname}}
      {\expandafter\uppercase\expandafter{\csname #2section\endcsname}}

    \def\efloat@type{#2}%
    \processdelayedfloat@hook
    \@nameuse{process#2s@hook}%
    \clearpage
    \@input{\jobname.#1}%
  }{}}
   
   }{}%
}%

  \makeatother

\newcommand{\diff}{{\rm d}}
\newcommand{\dln}{{\rm dln}}
\newcommand{\Ms}{{\ensuremath{\mathrm{M}_{\odot}}}}
\newcommand{\Rs}{{\ensuremath{\mathrm{R}_{\odot}}}}

\newcommand{\Mpy}{\Ms\,{\rm yr}{\ensuremath{^{-1}}}}
\newcommand{\dm}{\ensuremath{\dot M}}
\newcommand{\Edd}{{\ensuremath{\Gamma}}}

\newcommand{\vit}{{\ensuremath{\rm v}}}
\newcommand{\vkep}{{\ensuremath{\vit_{\rm Kep}}}}

\newcommand{\vcrit}{{\ensuremath{\vit_{\rm crit,1}}}}
\newcommand{\vvcrit}{{\ensuremath{\vit^2_{\rm crit,1}}}}
\newcommand{\vog}{{\ensuremath{\vit_{\rm crit,2}}}}
\newcommand{\vvog}{{\ensuremath{\vit^2_{\rm crit,2}}}}

\newcommand{\Rec}{{\ensuremath{R_{\rm eq,crit}}}}

\newcommand{\RRec}{{\ensuremath{R^2_{\rm eq,crit}}}}
\newcommand{\prad}{{\ensuremath{P_{\rm rad}}}}

\def\Lya{Ly$\alpha$}
\def\spose#1{\hbox to 0pt{#1\hss}}
\def\lta{\mathrel{\spose{\lower 3pt\hbox{$\mathchar"218$}}
     \raise 2.0pt\hbox{$\mathchar"13C$}}}
\def\gta{\mathrel{\spose{\lower 3pt\hbox{$\mathchar"218$}}
     \raise 2.0pt\hbox{$\mathchar"13E$}}}

\def\HI{\hbox{\rm H~$\scriptstyle\rm I$}}
\def\HII{\hbox{\rm H~$\scriptstyle\rm II$}}
\def\nH{{\rm H}}
\def\nHI{{\rm HI}}
\def\nHII{{\rm HII}}

\newcommand{\msunyr}{{\ensuremath{\mathrm{M}_{\odot}}}~{\rm yr}^{-1}}
\newcommand{\mdot}{\dot{M}_*}

\begin{document}

\title{~\mbox{Formation of the first stars and black holes}}
\author{L.~Haemmerl\'e, L.~Mayer, R.~S.~Klessen, T.~Hosokawa, P.~Madau \& V.~Bromm}
{\raggedright }\institute{}\titlerunning{{\mbox{}}}

\authorrunning{Haemmerl\'e, Mayer, Klessen, Hosokawa, Madau \& Bromm}
        \date{Received: date / Accepted: date}

      \maketitle 
      \keywords{}

\begin{abstract}

We review the current status of knowledge concerning the early
phases of star formation during cosmic dawn. This includes the first
generations of stars forming in the lowest mass dark matter halos in which
cooling and condensation of gas with primordial composition is possible at
very high redshift ($z > 20$), namely metal-free Population III stars, 
and the first generation of massive black holes
forming at such early epochs, the so-called black hole seeds. The formation of
black hole seeds as end states of the collapse of Population III stars, or via 
direct collapse scenarios, is discussed. In particular, special emphasis is
given to the physics of supermassive stars as potential precursors of direct
collapse black holes, in light of recent results of stellar evolution models,
and of numerical simulations of the early stages of galaxy formation.
Furthermore, we discuss the role of the cosmic radiation produced by the early generation of stars and black holes at high redshift in
the process of reionization.

\end{abstract}


\section{First stars}
\label{sec:popiii}

The first generation of stars, the so-called Population III (or Pop. III) built up from truly metal-free primordial gas. They have long been thought to live short, solitary lives, with only one extremely massive star forming in each dark matter halo with about 100 solar masses or more  \citep{Omukai2001,Abel2002,Bromm2002,Oshea2007}. The idea was that first star formation is simple and one only needs to know the initial Gaussian density perturbations of material at very high-redshift which are very well understood, e.g.\ from measuring cosmic microwave background (CMB) anisotropies \citep{Planck2016}, the growth of cosmological structures, and  the heating and cooling processes in the primordial gas. However, this simple picture has undergone substantial revision in recent years, and we now understand that stellar birth in the early Universe is subject to similar complexity as star formation at present days. Numerical simulations indicate that fragmentation is a wide-spread phenomenon in first star formation \citep{Clark2011b, Greif12}, and consequently that Pop. III stars form as members of multiple stellar systems with separations as small as the distance between the Earth and the Sun \cite[e.g.][]{Turk09, Clark11, Greif2011, Smith11, Stacy2013}. Studies that do include radiative feedback \citep{Hirano2014, Hirano2015, Hosokawa2016}, magnetic fields \citep{Machida2006, Machida2008, Schleicher2009, Schliecher2010dyn, Sur2010, Sur2012, Turk2012, Schobera, Schoberb, Bovino2013NJP}, dark matter annihilation \citep{Smith2012, Stacy2014}, as well as the primordial streaming velocities \citep{Tseliakhovich2010, Greif2011, Maio11, Stacy2011} add further levels of complexity. It turns out that all of these processes are relevant and need to be incorporated in our theoretical models. There is agreement now that primordial star formation is equally dynamic and difficult to understand as stellar birth at present days \cite[e.g.][]{Maclow2004, Mckee2007, Klessen2016}
 
In this section we investigate how star formation progresses once the combination of gravitational collapse and cooling leads to a strong accretion flow into the center of a high-redshift halo, and we discuss the most important physical processes that govern stellar birth on small scales. We focus on the standard pathway to Pop. III star formation. We argue that the accretion disk that forms around the first object is likely to fragment, which typically results in the formation of a cluster of stars with a wide range of masses rather than the built-up of a single high-mass object. We then speculate about how stellar feedback, the presence of magnetic fields, the potential energy input from dark matter annihilation and the possible existence of large-scale streaming velocities between baryons and dark matter may influence this picture. Supermassive stars and the possible seeds for supermassive black holes will be discussed in Sections \ref{sec:SMS} and \ref{sec:MBHS}, respectively. 
The implications on the reionization of the Universe are addressed in Sect.~\ref{sec:reionization}

\subsection{Spherical collapse models}
\label{subsec:1D-collapse}
\begin{figure}[th]
\begin{center}
\includegraphics[width=0.85\textwidth]{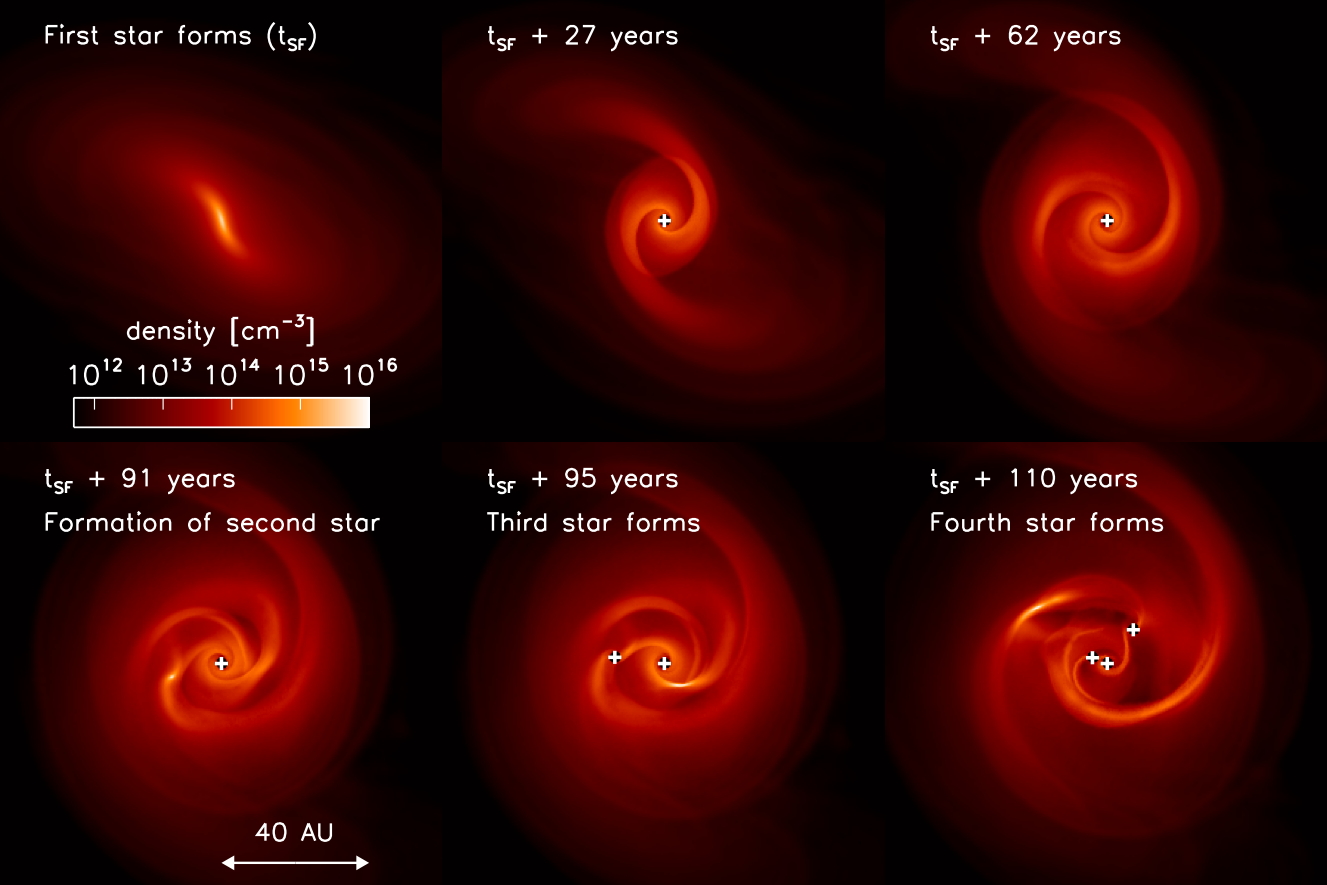}
\end{center}
 \caption{Density evolution in a 120 AU region around the first protostar, showing the build-up of the protostellar disk and its eventual fragmentation. The prominent two-arm spiral structure is caused by the gravitational instability in the disk, and the resulting gravitational torques provide the main source of angular momentum transport that allows disk material to accrete onto the protostar. Eventually, as mass continues to pour onto the disk from the infalling envelope, the disk becomes so unstable that regions in the spiral arms become self-gravitating in their own right. The disk fragments and a multiple system is formed.  Figure adopted from \citet{Clark11}.
 }
\label{fig:Clark-1}
\end{figure}

The  simplest model for primordial star formation is the run-away collapse of a spherically symmetric halo with mass exceeding the critical value for the cooling instability to set in \citep{Glover2005,Glover2013}. This was studied extensively in the early 2000's \citep{Abel2002,Bromm2002,Bromm04, Yoshida2003, Yoshida2006, Yoshida2008}. Figure \ref{fig:Clark-1} adopted from \citet{Clark11} shows a typical example of such a system. And from the numbers provided, we can obtain a crude estimate of the accretion rate onto the center of the halo as 
\begin{equation}
\dot{M} =\, \zeta \dfrac{M_{\rm J}}{\tau_{\rm ff}} \propto \dfrac{c_{\rm s}^{3}}{G}\;,
\label{eq:accretion-rate}
\end{equation}
where $M_{\rm J} \approx 50 \,{\rm M}_{\odot} \;\mu^{-2} n^{-1/2} T^{3/2}$ is the Jeans mass with number density $n$ measured in particles per cubic centimeter, temperature $T$ in Kelvin, and mean molecular weight $\mu$ being $\sim 1.22$ for primordial atomic and $\sim2.33$ for primordial molecular gas,  where $\tau_{\rm ff} \approx 52 \,\rm{Myr}\, \mu^{-1/2} n^{-1/2}$ is the free-fall time, and where $c_{\rm s} = (k_{\rm B}T/\mu m_{\rm H})^{1/2}$ is the sound speed with the Boltzmann constant $k_{\rm B}$ and the proton mass $m_{\rm H}$. Because of $\dot{M} \propto T^3$, warmer gas leads to higher accretion rates. The factor  $\zeta$ can take values of up to several tens depending on the actual density profile and by how much the actuall gas mass $M$ exceeds the Jeans mass $M_{\rm J}$ \cite[for classical studies, see][]{Shu1977,Larson1969, Penston1969, Whitworth1985}. With a typical gas temperature of  $T\sim 1500\,$K we get accretion rates of $\dot{M} \approx 2 \times 10^{-3}\,$M$_{\odot}\,$yr$^{-1}$.

\subsection{Disk fragmentation}
\label{subsec:fragmentation}
These early calculations typically stopped when the object in the center reached number densities of $n \approx 10^{16}\,$cm$^{-3}$ or more, because then the computational timestep became prohibitively small.  At this time the protostar has a mass of only $\sim 10^{-3}\,$M$_{\odot}$. However, as more and more gas falls to the center it also carries along more and more angular momentum. Eventually this gas settles into a rotationally supported disk rather than being consumed by the central protostar. Theoretical and numerical models that follow the build-up and longer-term evolution of this disk show that the system is prone to fragmentation. 

Disk fragmentation on various spatial and temporal scales is reported by various authors \citep{Machida2008, Clark11, Greif2011,Greif12, Dopcke2013,Susa13,Susa14,Hosokawa2016,Stacy2013,Stacy2016,Turk09, Wollenberg2020}. Under typical conditions of Pop. III star formation the mass load onto the disk by accretion from the infalling envelope exceeds its capability to transport this material inwards by gravitational or magnetoviscous torques, that is by spiral arms \citep{Binney1987} or by the magnetorotational instability \citep{Balbus1998}. Consequently the disk becomes Toomre unstable and fragments. This preferentially occurs at the outer edge of the disk. The instability region moves outwards as the disk grows larger by accretion of higher angular momentum material, and the formation of new protostars occurs at larger and larger radii as the evolution progresses. This is indicated in Figure \ref{fig:Clark-1}, adopted  from \citet{Clark11}, which shows the build-up of altogether four protostars within only about hundred years after the formation of the first object. 

Disk fragmentation has important consequences for the resulting stellar mass spectrum. As matter flows through the disk towards the center, it first encounters the Hill volume of secondary protostars further out and preferentially gets swallowed by these objects. Otherwise, this gas would have continued to move inwards and would eventually be accreted by the central object. Clearly, disk fragmentation limits the mass growth of the primary star in the center, and so this  process has been termed 'fragmentation-induced starvation' in the context of present-day star formation \citep{Kratter2006,Peters2010, Girichidis2011, Girichidis2012,Girichidis2012B}. Some of these protostars may get ejected by dynamical encounters with other protostars or fragments while some may move inwards to get accreted by the central object \citep{Clark2011b, Greif12, Smith2012, Stacy2016}. As a consequence of these stochastic behavior, the mass spectrum of Pop. III stars is expected be very broad, possibly reaching down into the substellar regime. We conclude that the standard pathway of Pop. III star formation leads to a stellar cluster with a wide distribution of masses rather than the build-up of one single high-mass object.


\subsection{Radiative feedback}
\label{subsec:radiation}

Massive Pop III stars are expected to enter the hydrogen burning main sequence while still accreting \citep{Zinnecker2007,Maeder2009}. For accretion rates below $\dot{M} \approx 10^{-3}\,$M$_{\odot}$ they are thought to be compact and very hot at their surface \citep{Hosokawa2009, Hosokawa2010, Hosokawa2012}, implying that they  emit large numbers of ionizing photons \citep{Schaerer2002} that can significantly influence their birth environment. For rates much larger than $\dot{M} \gtrsim 10^{-2}\,$M$_{\odot}$ stellar evolution calculations suggest that stars remains bloated and fluffy \citep{Hosokawa13, Umeda2016, Woods2017,haemmerle2018a}. In this case, the surface temperature is relatively low, and although very luminous, these stars do not emit much ionizing radiation and could be able to maintain this high accretion flow for a very long time. In the right environment this could possibly lead to the formation of supermassive stars as we discuss in Section \ref{sec:SMS}.

Compact and hot Pop. III stars create HII regions which eventually break out of the parental halo and may affect stellar birth in neighboring systems \cite[see, e.g.][]{Kitayama2004,Whalen2004, Alvarez2006, Abel2007, Yoshida2007, Greif2008, Wise2012, Wise2012b, Jeon2014}. The question is when this happens and how radiative feedback influences the immediate birth environment. When addressing this problem, we can seek guidance from models of stellar birth at present days. Recent simulations  \citep{Yorke2002,Krumholz2009, Kuiper2010, Kuiper2011, Peters2010, Peters2011, Commercon2011, Rosen2016} indicate that once a protostellar accretion disk has formed, it quickly becomes gravitationally unstable and so material in the disk midplane flows inwards along dense filaments, whereas  radiation escapes through optically thin channels above the disk. Even ionized material can be accreted, if the accretion flow is strong enough \citep{keto2007, Peters2010}. Radiative feedback is found not to be able to shut off the accretion flow, instead it is the dynamical evolution of the disk material that controls the mass growth of individual protostars. 

Due to the lack of metals and dust, protostellar accretion disks around Pop. III stars can cool less efficiently and are much hotter than disks at present days \citep{Tan2004}. Furthermore, the stellar radiation field couples less efficiently to the  surrounding because the opacities are smaller. It is thus not clear how well the above results can be transferred to the primordial case. Numerical simulations in two dimensions by \citet{Hosokawa11}, \citet{Hirano2014} and \citet{Hirano2015}  find that radiative feedback can indeed stop stellar mass growth and quickly remove the remaining accretion disk, resulting in final stellar masses from a few $10\,$M$_\odot$ up to about $1000\,$M$_\odot$. However, these calculations cannot capture disk fragmentation and the formation of multiple stellar systems. Three-dimensional calculations by \citet{Stacy2012}, \citeauthor{Susa13} (\citeyear{Susa13,Susa14}), or \citet{Hosokawa2016} on the other hand find widespread fragmentation with a wide range of stellar masses down to $\sim 1\,$M$_\odot$. All calculations have limitations and shortcomings, either in terms of resolution or in the number of physical processes included. High resolution is essential for correctly computing the recombination rate in ionized or partially ionized gas. It goes quadratic with the density, and consequently, not resolving the high-density shell of gas swept up by the expanding ionized gas can result in incorrect HII region expansion rates and sizes \cite[see, e.g.][]{Rahner2017, Rahner2019}. By a similar token, treating only ionizing radiation and neglecting the impact of Lyman-Werner wavebands can lead to large errors in the overall photon escape fraction \citep{Schauer2015}. Consequently, any firm conclusions about the resulting mass spectrum of Pop. III stars in the presence of radiative feedback seems premature at this stage.

\subsection{Magnetic fields}
\label{subsec:magnetic-fields}

Also the presence of dynamically important magnetic fields will influence disk evolution and fragmentation. Unfortunately, we know very little about magnetic fields in the high-redshift universe. Theoretical models predict that magnetic fields could be produced in various ways, for example via the Biermann battery \citep{Biermann1950}, the Weibel instability \citep{Lazar2009, Medvedev2004}, or thermal plasma fluctuations \citep{Schlickeiser2003}. Other models invoke phase transitions that occur during cosmic inflation \citep{Sigl1997, Grasso2001,Banerjee2003, Widrow2012}. In each case, the inferred field strengths are very small. The only way to obtain dynamically significant magnetic fields is through the small-scale turbulent dynamo, which is very efficient in amplifying even extremely small primordial seed fields to the saturation level. This process is very fast and acts on timescales much shorter than the dynamical free-fall time. An analytic treatment is possible in terms of the Kazantsev model \citep{Kazantsev1968, Subramanian1998, Schobera,Schoberb}. Once backreactions become important, the growth rate slows down, and saturation is reached within a few large-scale eddy-turnover times \citep{Scheko02, Schekochihin2004, Schober2015}. 

Depending on the properties of the turbulent flow, the magnetic energy density at saturation is thought to lie between 0.1\% and a few 10\% of the kinetic energy density \citep{Federrath2011, Federrath2014}. Fields this strong affect the evolution of protostellar accretion disks: They remove angular momentum from the star-forming gas \citep{Machida2008,Machida2013, Bovino2013NJP, Latif2013a,LatifMag2014}, drive protostellar jets and outflows \citep{Machida2006}, and they reduce the level of fragmentation in the disk \citep{Turk2011, Peters2014}. Altogether, we expect Pop. III clusters to have fewer stars with higher masses than predicted by purely hydrodynamic  simulations. However, a full assessment of these differences still needs to be worked out in detail.

\subsection{Streaming velocity}
\label{subsec:streaming}

Prior to recombination, baryons are tightly coupled to photons resulting in a standing acoustic wave pattern \citep{Sunyaev1970}. This results in oscillations between baryons and dark matter  with relative velocities of about $30\,$km$\,$s$^{-1}$ and coherence lengths of several $10\,$Mpc to $100\,$Mpc \citep{Silk1986} at $z \approx 1000$. After recombination, baryons  are no longer tied to photons, their sound speed drops to $\sim 6\,$km$\,$s$^{-1}$, and the velocity with respect to the dark matter component becomes supersonic with Mach numbers of ${\cal M} \approx 5$  \citep{Tseliakhovich2010}. In the subsequent cosmic evolution, the relative streaming velocity decays linearly with $z$, and reaches  $\sim 1\,$km$\,$s$^{-1}$ at the onset of first star formation at $z \approx 30$. Numerical simulations with streaming velocities suggest a reduction of  the baryon overdensity in low-mass halos, a delay of the onset of cooling, and a larger critical mass for collapse \citep{Greif2011, Stacy2011, Maio2011, Naoz2012, Naoz2013, Oleary2012, Latif2014Stream, Schauer17a}. This can have substantial impact on the resulting $21\,$cm emission \cite[][see also Section \ref{sec:reionization}]{Fialkov2012, McQuinn2012,Visbal2012}. Gas in halos which are located in regions of large streaming velocities will be more turbulent than in more quiescent systems, and so we expect more fragmentation and a bias towards smaller stellar masses \citep{Clark2008}. However, this process has not yet been modeled with sufficient resolution, and so a reliable prediction of the stellar mass spectrum in regions with large streaming velocities is still outstanding.

\subsection{Dark matter annihilation}
\label{subsec:DM-annihilation}

Despite its importance for cosmic evolution and structure formation, the true physical nature of dark matter is unknown and so the studies of the impact of dark matter annihilation on first star formation are still highly speculative. As baryons collapse in the centers of primordial halos, they may drag along dark matter particles in a processes termed 'adiabatic contraction' \citep{Blumenthal1986}. It can lead to a density increase of several orders of magnitude, and as the annihilation rate scales quadratically with the density the corresponding energy input and ionization rate may become large enough to influence the dynamics of the gas. \citet{Spolyar2008} and \citet{Freese2009} suggest that this process could potentially overcome the cooling provided by H$_2$ and eventually halt further collapse. They call the resulting object `dark star'. It has a size of few astronomical units and is powered by dark matter  annihilation rather than by nuclear fusion. If dark matter particles also scatter weakly on baryons this could lead to a structure that is stable for a long time \citep{Freese2008, Iocco2008A,Iocco2008B, Yoon2008,Hirano2011}. 

There are several concerns that have been raised in this context. First, it is not clear whether collapse stalls once the energy input from dark matter annihilation becomes comparable to the cooling rate, because the larger heating rate may catalyze further H$_2$ formation and lead to a correspondingly larger cooling rate \citep{Ripamonti2010}. Second, the underlying assumption of perfect alignment between dark matter cusp and gas collapse is most likely violated in realistic star formation conditions. For example,  \citet{Stacy2012, Stacy2014} find that non-axisymmetric perturbations lead to a separation between dark matter cusp and collapsing gas and causes the annihilation energy input to be too small for dark star formation. Similarly, \citet{Smith2012} see no evidence for the build-up of dark stars in their simulations.  However, they find that dark matter annihilation has the potential to modify the dynamics of the accretion disk and reduce the level of disk fragmentation. As before, more work is required before we are able to draw firm conclusions.


\section{Supermassive stars}
\label{sec:SMS}

Motivated by recent discoveries of supermassive black holes (SMBHs) exceeding $10^9~\Ms$ at high redshifts $z \gtrsim 6$, the formation of supermassive stars (SMSs) in the early universe has drawn great attention for recent years. A possible formation pathway of the SMSs is the direct collapse (DC) model \citep[e.g.,][]{BL03}, where the SMSs gain the huge mass via very rapid mass accretion, typically 0.1 -- 10 \Mpy\ expected in atomic cooling primordial haloes.


Although the evolution and final fates of the SMSs had been investigated in the literature for past decades
\citep[e.g.,][]{Hoyle1963,Osaki66,Unno71,Fricke73,Fuller86},
the actual formation process had been ignored in such previous studies (`monolithic' models); a SMS in the stable nuclear burning phase was assumed initially, and then the subsequent evolution was followed. However, considering the formation process is critically important. For instance, the stellar radiative feedback against the accretion flow is a potential obstacle in forming massive stars. Even in the normal primordial star formation, where the typical stellar mass is thought to be $\sim 100~\Ms$, the stellar UV feedback may be strong enough to halt the mass accretion onto the stars \citep[e.g.,][]{MT08,HOYY11,Sugimura20}. In the DC model, where the stellar mass is supposed to become much higher, the more tremendous UV radiative feedback may prevent the SMS formation early on. Solving the evolution of accreting SMSs from their birth stage is necessary to evaluate the strength of the UV feedback. In the same vein, the earlier evolution with the rapid mass accretion also affects the final fate of the SMS, i.e., at which point it collapses into a BH. This determines how massive SMBH seeds are finally provided in the DC scenario.

\subsection{Numerical Modeling of Accreting Stars}

Numerical modeling of accreting stars has been developed in some different contexts, including the binary evolution \citep[e.g.,][]{Kippenhahn77,Neo77}, present-day star formation \citep[e.g.,][]{SST80,Stahler88,Palla91,Behrend01,HO09,haemmerle2016}, and primordial star formation \citep[e.g.,][]{SPS86,OP03}. In parallel with the emergence of the DC model, the similar method has been applied for the evolution of SMSs with extremely high accretion rates exceeding $10^{-2}~\msunyr$.


Regardless of such different contexts, the above studies basically solve the same governing equations to construct stellar models, 
\begin{equation}
\left( \frac{\partial r}{\partial M} \right)_t = \frac{1}{4 \pi \rho r^2},
\label{eq:con} 
\end{equation}
\begin{equation}
\left( \frac{\partial P}{\partial M}  \right)_t = - \frac{GM}{4 \pi r^4}, 
\label{eq:mom}
\end{equation}
\begin{equation}
\left( \frac{\partial L}{\partial M} \right)_t 
= \epsilon - T \left( \frac{\partial s}{\partial t}  \right)_M ,
\label{eq:ene}
\end{equation}
\begin{equation}
\left( \frac{\partial s}{\partial M} \right)_t
= \frac{G M}{4 \pi r^4} \left( \frac{\partial s}{\partial p} \right)_T
  \left( \frac{L}{L_s} - 1  \right) C ,
\label{eq:heat}
\end{equation}                                                                
where $M$ is the mass coordinate, $\epsilon$ the energy production rate via nuclear fusion, $s$ the specific entropy, $L_s$ the radiative luminosity with adiabatic temperature gradient, $C$ the coefficient determined by the mixing-length theory when $L > L_s$ ($C=1$ otherwise), and the remaining quantities have ordinary meanings. Effects of the mass accretion are incorporated in the term
$T (\partial s / \partial t)_M$ in Eq.(\ref{eq:ene}), which
is approximately discretized as
\begin{equation} 
\left( \frac{\partial s}{\partial t}  \right)_M = 
  \left( \frac{\partial s}{\partial t} \right)_m 
+ \dot{M}_\ast  \left( \frac{\partial s}{\partial m}  \right)_t ,
\label{eq:relm}
\end{equation}
where $m$ is the inverse mass coordinate measured from the surface, $m \equiv M_{\ast} - M$. An evolutionary calculation begins with an initial model with small mass, and advances with updating models by adding the accreting gas mass $\mdot \Delta t$, where $\Delta t$ is the time step. 

\subsubsection{Accretion-induced ``Inflation'' of Stars}

\begin{figure}
\centerline{\includegraphics[width=9.5cm]{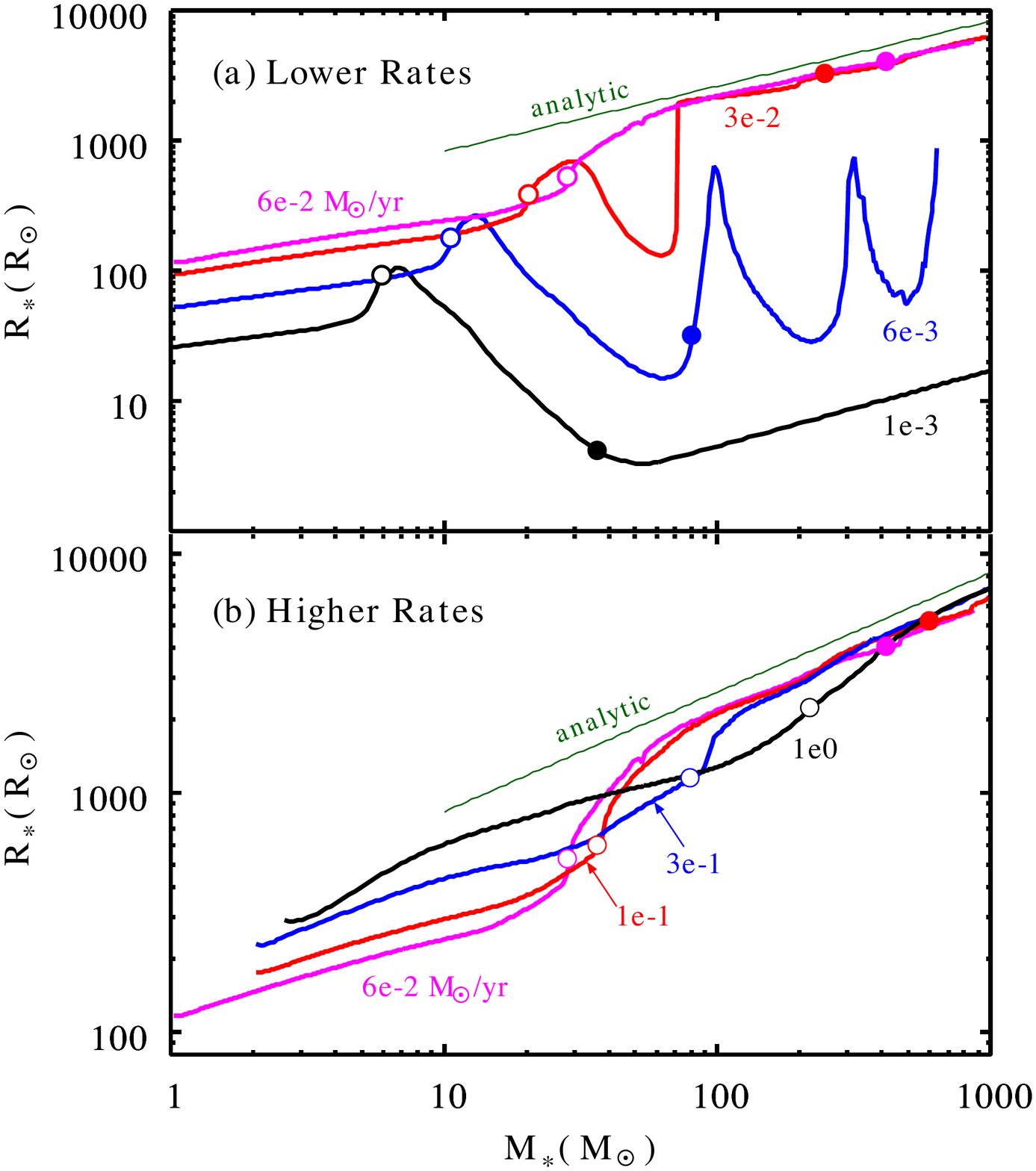}}
\caption{Evolution of the stellar radius with the different accretion rates, from \cite{HOY12}. The upper and lower panels present the evolution with the rates lower and higher than $6 \times 10^{-2}~\msunyr$. The different lines represent the accretion rates as indicated by the labels. 
The thin green line represents the analytic mass-radius relationship given by equation (\ref{ra_anal}). The filled circles on the lines denote the epoch when the hydrogen burning initiates near the center.} 
\label{ra_fig1}
\end{figure}

Figure~\ref{ra_fig1} summarizes the results by \cite{HOY12}, who have numerically followed the evolution of protostars accreting at various constant rates including high rates exceeding $10^{-2}~\msunyr$. The black curve in panel (a) represents the case with $\mdot = 10^{-3}~\msunyr$, a relatively low accretion rate typical for the normal primordial star formation. A striking feature for this case is the stellar contraction occurring for $6~\Ms \lesssim M_* \lesssim 30~\Ms$. This is the so-called Kelvin-Helmholtz (KH) contraction stage, during which the star loses its energy by radiating away.
The stellar effective temperature and emissivity of ionizing photons substantially rises during this, and the UV feedback against the accretion flows begins to operate. The presence of the KH contraction thus marks the onset the stellar UV feedback. Since the central temperature gradually rises, the hydrogen burning starts just after the stellar mass exceeds $30~\Ms$. After this point, the stellar radius turns to rise following the mass-radius relation for zero-age main-sequence (ZAMS) stars.


By contrast, the above evolution drastically changes with the more rapid accretion. Panel (a) shows that the KH contraction stage only spans the narrower mass range for such cases. 
The contraction turns to the expansion with the accretion rates $6 \times 10^{-3}~\msunyr$ and $0.03~\msunyr$. With the highest rate $0.06~\msunyr$, the contraction never occurs and the stellar radius monotonically increases as the star accretes the gas. The stellar radius finally reaches $\simeq 7000~\Rs$ (or nearly 30~AU) when the stellar mass is $\sim 1000~\Ms$. 
Panel (b) summarizes the evolution with the even higher accretion rates, $0.06~\msunyr \leq \mdot \leq 1~\msunyr$, showing further intriguing features, i.e., the curves corresponding to the different accretion rates represent the same mass-radius relation for $M_* \gtrsim 300~\Ms$.  The radius at $M_* \simeq 1000~\Ms$ is consequently $\simeq$ 30~AU for all the cases.


The above behavior is well interpreted with the following analytic argument. Let us begin with the general expression of the stellar luminosity
\begin{equation}
 L_* = 4 \pi R_*^2 \sigma T_{\rm eff}^4,
\label{ra_eqlum}
\end{equation}
where $\sigma$ is the Stefan-Boltzmann constant.
As for very massive stars, the luminosity is very close to the Eddington value,
\begin{equation}
 L_* \simeq L_{\rm Edd} \propto M_* .
\label{ra_eqledd}
\end{equation}
In the current cases with the high accretion rates, a stellar surface layer inflates because it always traps a part of the heat released from the contracting interior. The effective temperature accordingly decreases to several $\times$ 1000~K, where H$^-$ bound-free absorption mostly contributes the opacity. Since the H$^-$ opacity has a very strong temperature-dependence, the effective temperature is almost fixed at a constant value. Substituting equation (\ref{ra_eqledd}) and $T_{\rm eff} \simeq 5000$~K into equation (\ref{ra_eqlum}), we derive the mass-radius relation
\begin{equation}
R_* \simeq 2.6 \times 10^3~R_\odot 
\left( \frac{M_*}{100~M_\odot} \right)^{1/2} ,
\label{ra_anal}
\end{equation}
which is indicated by the thin green lines in Figure~\ref{ra_fig1}. The above derivation includes no dependence on the accretion rate, which explains why the different lines in Figure~\ref{ra_fig1} eventually converge into the same mass-radius relation.


\subsubsection{Birth of Accreting SMSs}

\begin{figure}
\centerline{\includegraphics[width=9.5cm]{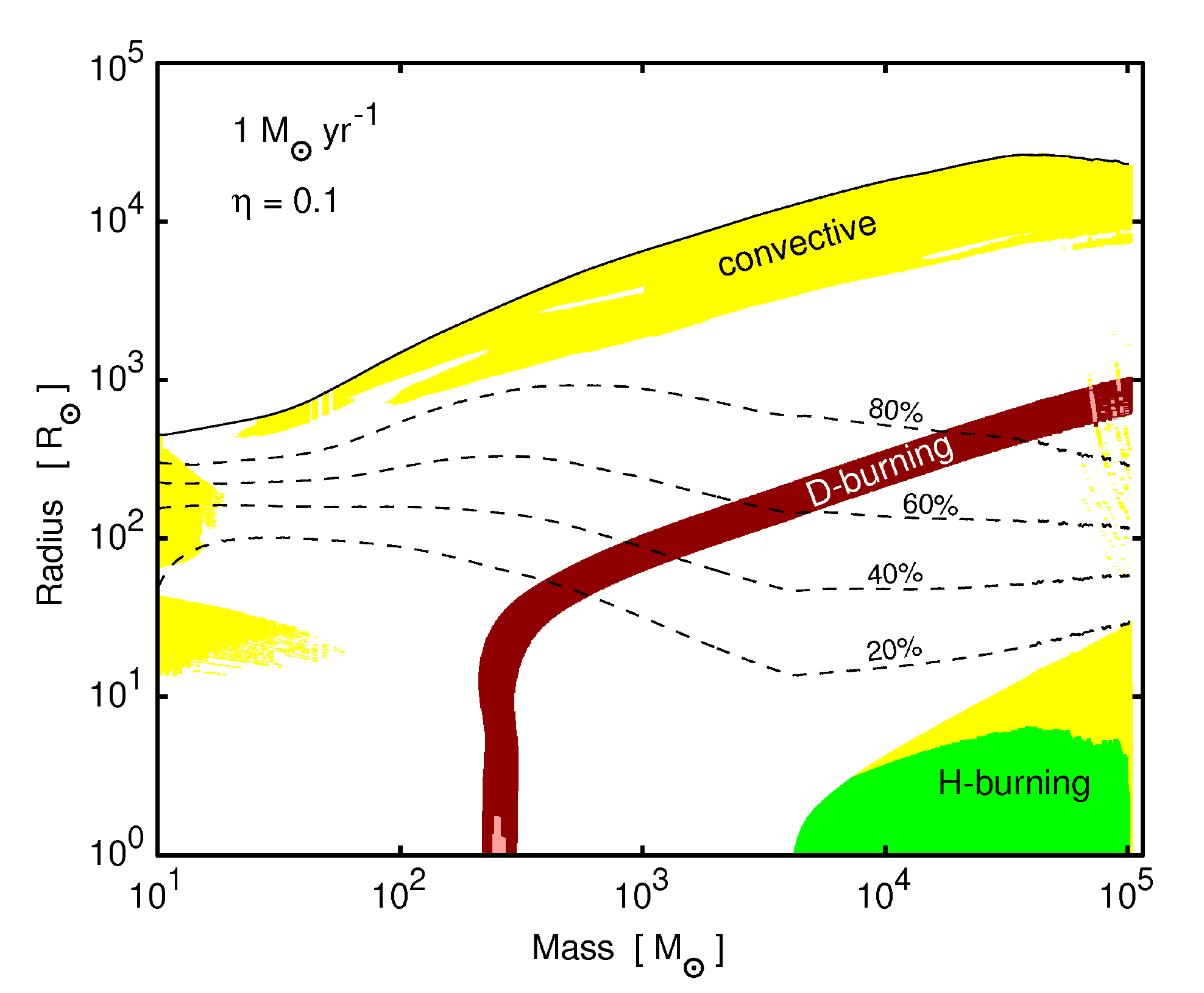}}
\caption{Evolution of the stellar interior structure until the stellar mass reaches $10^5~\Ms$ with the accretion rate $\mdot = 1~\msunyr$, from \cite{Hosokawa13}.
The black solid and dashed lines represent the radial positions of the stellar surface and mass coordinates enclosing $80\%$, $60\%$, $40\%$, and $20\%$ of the total mass in descending order. The white and yellow areas represent radiative and convective zones where no nuclear fusion takes place. The brown stripe and pink zones indicate the radiative and convective layers where deuterium burning occurs. The green area represents a convective core where hydrogen burning occurs. } 
\label{ra_fig2}
\end{figure}

\citet{Hosokawa13} have further followed the later evolution until the stellar mass reaches $10^4 \sim 10^5~\Ms$. Figure~\ref{ra_fig2} illustrates the evolution of the stellar interior structure with the constant accretion rate $1~\msunyr$. 
We see that the monotonic increase of the stellar radius continues after the stellar mass exceeds $1000~\Ms$. The analytic formula given by equation (\ref{ra_anal}) still provides a good fit for the evolution of the stellar radius, though it only slightly overestimates the radius for $M_* \gtrsim 3 \times 10^4~\Ms$. The similar evolution has been reported by the different authors~\citep[e.g.,][]{Umeda2016, haemmerle2018a}. 


Figure~\ref{ra_fig2} shows that a surface layer which only has a small fraction of the total mass significantly inflates to cover a large part of the stellar radial extent. The star has a bloated envelope and compact core like red giants. Most of the stellar mass is contained in the central core, which gradually contracts as the star accretes the gas. The central temperature rises during this, as in the ordinary KH contraction. As a result, the nuclear burning starts just after the total stellar mass exceeds $\simeq 4000~\Ms$. The central convective core gradually extends after that, owing to the heat input caused by the nuclear fusion. The outer convective layer remains bloated as long as the rapid mass accretion continues, so that the stellar radius is much larger than the ZAMS stars with the same masses.


As described above, SMSs under the very rapid mass accretion greatly inflate. Their effective temperatures are only $\simeq 5000$~K, and the UV emissivity is much smaller than that of non-accreting stars. The resulting UV feedback is thus too weak to disturb the accretion flow toward the stars, so that the rapid growth of the stellar mass should continue as far as the high accretion rates exceeding $\sim 10^{-2}~\msunyr$ are maintained. In the standard DC model, SMSs exceeding $10^5~\Ms$ may appear in less 0.1~Myr since the birth of the initial embryonic protostar, and they provides massive SMBH seeds once the gravitational collapse occurs via the general relativistic instability (Sect.~\ref{sec:SMSfin}). 

\subsubsection{Accretion at $\dm>10$ \Mpy}
\label{sec:SMSmax}

Beside the usual DC scenario based on atomic cooling haloes,
an alternatice formation route for massive BH seeds has been proposed,
relying on the merger of massive, gas-rich galaxies \citep{mayer2010,mayer2015,mayer2019},
as we describe in more details in Sect.~\ref{sec:MBHS}.
In this scenario, inflows as high as $10^5$ \Mpy\ are found on subparsec scales.

The evolution of SMSs under accretion at $\dm>10$ \Mpy, up to 1000 \Mpy, has been simulated in \cite{haemmerle2019c}.
The main qualitative difference compared to models at lower rates is that for such rates the accretion timescale $M/\dm$
becomes similar to or shorter than the dynamical time of the star.
Thus, above a limit in the accretion rate, the evolution proceeds hydrodynamically.
This limit is an increasing function of the stellar mass $M$,
since the perturbation to equilibrium induced by a given amount of mass added at the surface
becomes relatively weaker as the stellar mass grows.

\begin{figure}
\includegraphics[width=0.66\textwidth]{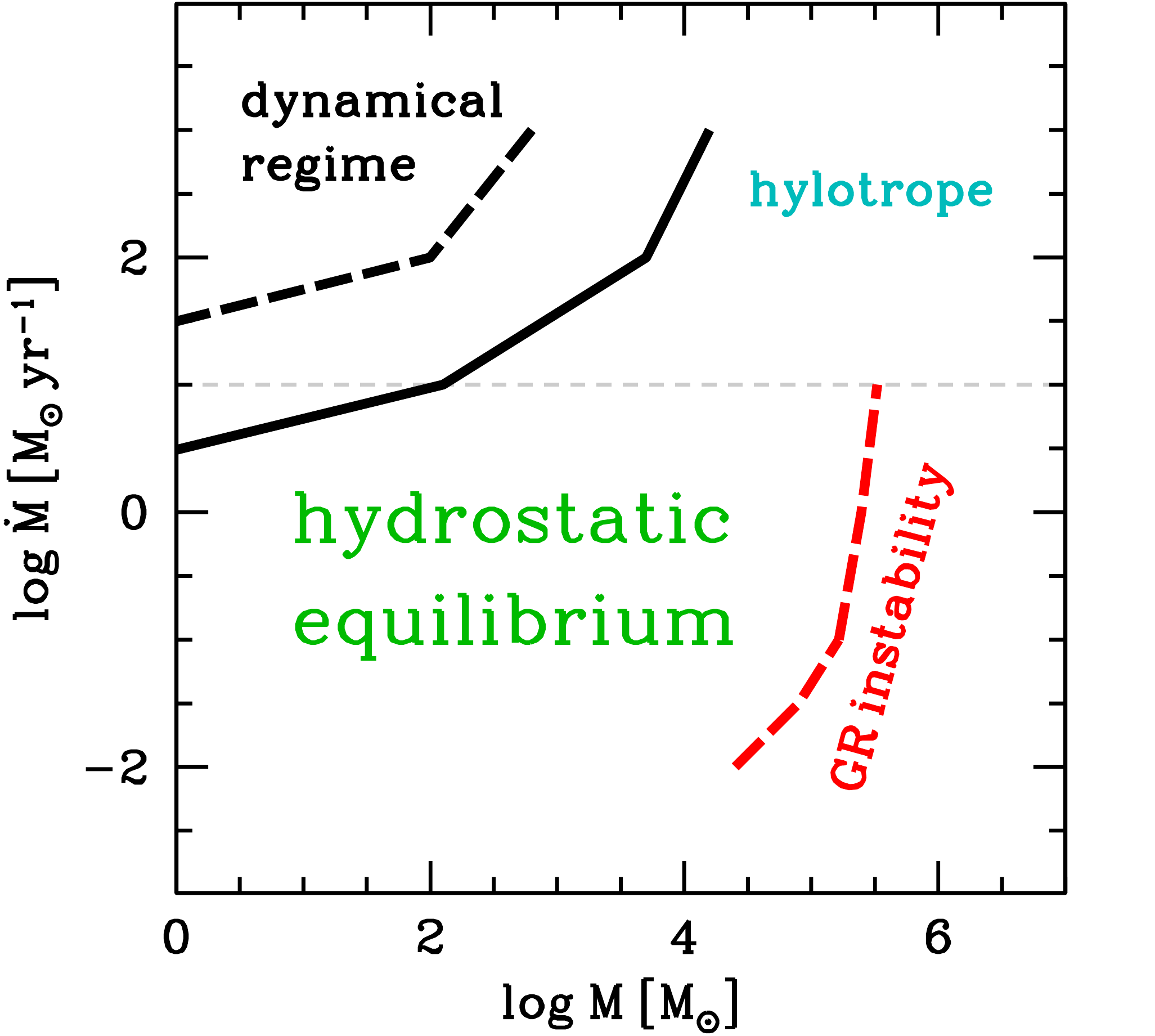}\label{fig-mmdot}
\caption{Limits to hydrostatic equilibrium in the $M-\dm$ plan, from \cite{haemmerle2019c}.
Above the solid black line, accretion is too rapid for the star to evolve in quasi-static equilibrium.
At the right-hand side of the red line, stars are GR unstable \citep{Woods2017}, see Sect~\ref{sec:SMSfin}.}
\end{figure}

The curve of $\dm(M)$ is shown on Fig.~\ref{fig-mmdot}.
Rates of $\gtrsim100$ \Mpy\ become consistent with equilibrium only when the accretor is supermassive
($M\gtrsim10^4$ \Ms).
If the conditions for equilibrium are satisfied, the star evolves in a qualitatively similar way as for lower rates:
the envelope is inflated and the evolutionary track follows the Hayashi limit upwards.
Most of the star is radiative, and hydrogen burns in the centre,
driving convection in a core that contains a few percent of the total stellar mass.
For $\dm=1000$~\Mpy, the mass fraction of the convective core can be as low as 1\% at the endpoint of the evolution.

If lower mass objects accrete at rates $\gtrsim100$ \Mpy, their evolution proceeds dynamically,
since the pressure and temperature gradients required for equilibrium cannot build.
The further evolution in the dynamical regime is not known.
In particular, the possibility for {\it dark collapse} \citep{mayer2019} shall be addressed with hydrodynamical simulations.

\subsubsection{The final collapse of SMSs}
\label{sec:SMSfin}

SMSs end their life through the general relativistic (GR) instability,
which is a radial pulsation instability that reflects photon's contribution to gravity \citep{chandrasekhar1964}.
GR effects impact negligibly the properties of stable SMSs, but they are critical for their stability,
since these stars are always close to the Eddington limit.
For $M>10^4$ \Ms, the ratio $\beta$ of the gas pressure to the total pressure is of the order of a few percent.
The first adiabatic exponent can be developed as
\begin{equation}
\Gamma_1=\left.{\dln P\over\dln\rho}\right\vert_{\rm ad}={4\over3}+{\beta\over6}+\mathcal{O}(\beta^2).
\end{equation}
Thus, $\Gamma_1$ exceeds its classical critical value $\Gamma_{\rm crit}=4/3$ by typically a percent.
In these conditions, the small corrections to $\Gamma_{\rm crit}$ arising from GR effects can trigger the collapse.
For polytropic models, the first order post-Newtonian correction to $\Gamma_{\rm crit}$ is
\begin{equation}
\Gamma_{\rm crit}={4\over3}+K{r_S\over r}+\mathcal{O}\left({r_S^2\over r^2}\right).
\end{equation}
where $r_S=2GM_r/c^2$ is the local Schwarzschild radius
and $K$ is a constant that takes a value of 1.12 for a polytrope $n=3$,
i.e. for an isentropic star in the limit $\beta\rightarrow0$.
As one considers increasing masses, the gas correction to $\Gamma_1$ decreases
while the GR correction to $\Gamma_{\rm crit}$ increases, until
\begin{equation}
{\beta\over6}<K{r_S\over r}	\qquad\Longrightarrow\qquad		\Gamma_1<\Gamma_{\rm crit}
\end{equation}
and the star becomes GR unstable.

\begin{figure}
\includegraphics[width=0.66\textwidth]{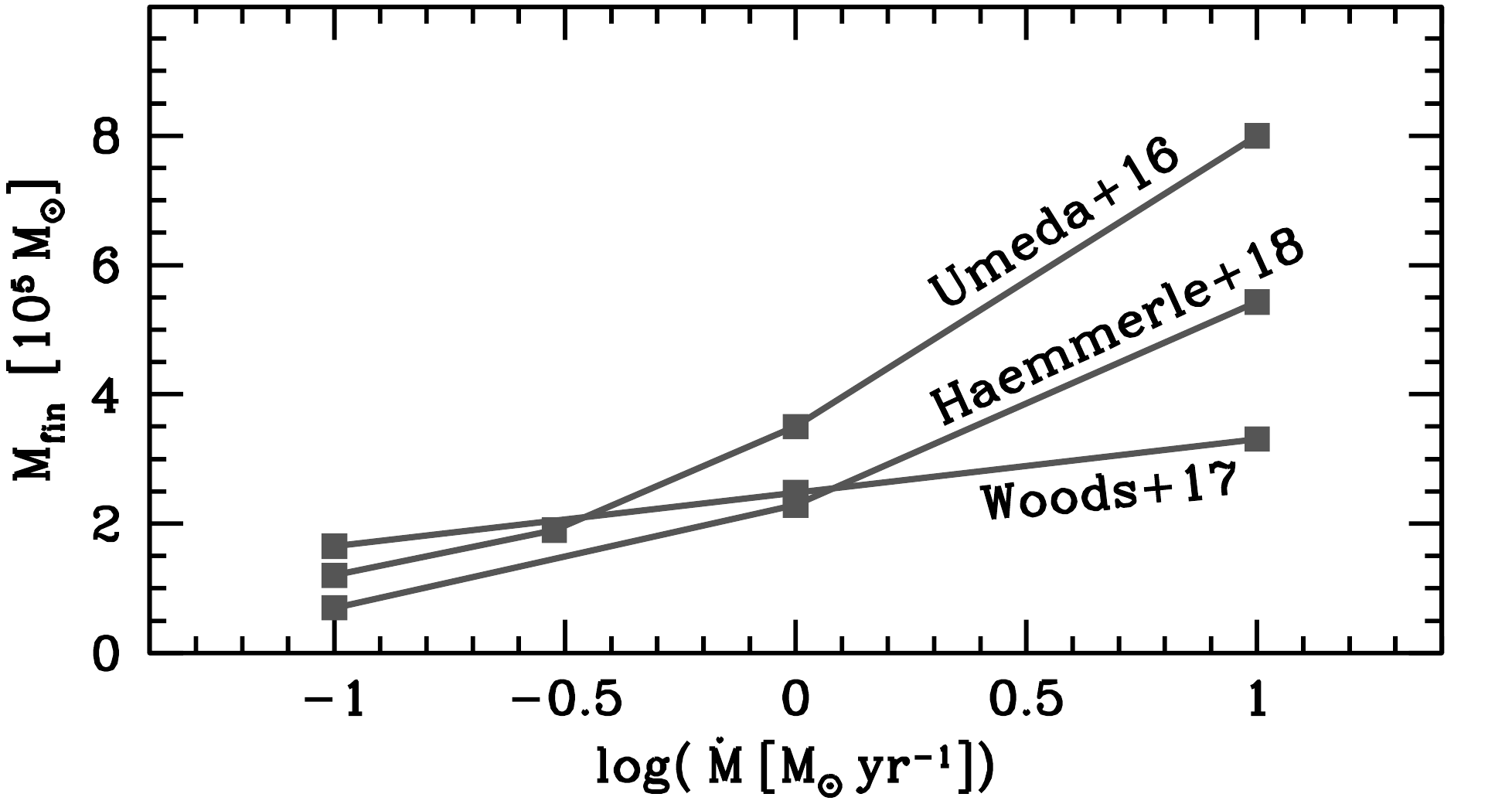}\label{fig-fin}
\caption{Mass-limits for GR stability of accreting SMSs, as a function of the (constant) accretion rate,
by \cite{Umeda2016}, \cite{Woods2017} and \cite{haemmerle2018a}.}
\end{figure}

The limit in mass above which accreting SMSs become GR unstable has been investigated in numerical simulations
\citep{Umeda2016,Woods2017,haemmerle2018a}, and are shown on Fig.~\ref{fig-fin}.
Discrepancies appear between the various studies (see \citealt{woods2019} for a detailed discussion),
but all the models indicate a limit of several $10^5$ \Ms\ for a range in accretion rate that covers two orders of magnitudes.
Models accreting at $100-1000$ \Mpy\ suggest that this limit remains always below $10^6$ \Ms\ even for such extreme rates \citep{haemmerle2019c}.

Hydrodynamical simulations of the final collapse of SMSs
indicate that the bulk of the stellar mass at the onset of instability forms a black hole in a dynamical time
\citep{saijo2002,saijo2004,liu2007,shibata2016b,sun2017,sun2018,li2018,uchida2017}.
In spherical symmetry, the whole mass is swallowed by the black hole,
but rotation allows for a few percent of the stellar mass to orbit outside the horizon as a torus.
Such asymmetries could trigger detectable gravitational wave (GW) emission.
In the presence of a magnetic field, relativistic jets are found to form,
which could trigger ultra-long gamma-ray bursts (ULGRBs).
Thus the final collapse of SMSs might allow for multi-messenger signatures of DC.
Thermonuclear explosion could occur in the presence of metals \citep{Fuller86,montero2012},
or in a narrow mass-range around 55 000 \Ms\ \citep{chen2014a}.

\subsection{Rotation in SMSs}
\label{sec:SMSom}

Rotation is key in the evolution of SMSs, for positive and negative reasons.
The negative reasons rely on the angular momentum problem, which is general to star formation:
the contraction from interstellar to stellar densities require mechanisms to extract angular momentum,
in order to avoid break-up by centrifugal forces \citep{spitzer1978,bodenheimer1995}.
This angular momentum barrier is particularly strong in massive star formation,
due to the importance of radiation pressure, the lack of magnetic fields, and the short evolutionary timescales
(e.g.~\citealt{lee2016,takahashi2017,haemmerle2017}).
It constitutes also a strong bottleneck for the efficient mass growth of SMBHs once formed \citep{sugimura2018}.
The positive reasons rely on the possibility for the final collapse of SMSs to trigger GWs and ULGRBs
(Sect.~\ref{sec:SMSfin}).
Rotation is required in both cases, in order to break spherical symmetry and to allow for the formation of collimated jets.

The effect of rotation on SMSs has been studied in details in the context of monolithic models,
i.e. neglecting accretion
\citep{fowler1966,bisnovatyi1967,appenzeller1971a,baumgarte1999a,baumgarte1999b,new2001,
shibata2016a,butler2018,dennison2019}.
In general, these studies assume the star is rotating as a solid body,
with a surface velocity corresponding to the Keplerian value.
The assumption of solid rotation relies on convection, which dominates in monolithic models.
The idea of Keplerian rotation follows from the argument of angular momentum conservation during the formation process.
As a consequence of Keplerian rotation, the star is expected to loose mass at the equator as it further contracts,
and this configuration is called the {\it mass-shedding limit} \citep{baumgarte1999a}.
The most striking result of these studies is the possibility for stars of $10^8-10^9$ \Ms\ to be stabilised by rotation against GR
\citep{fowler1966,bisnovatyi1967}.
Moreover, the rotational properties of monolithic SMSs at collapse are found to be universal,
in the sense that their spin parameter is
\begin{equation}
a=\frac{cJ}{GM^2}=0.97,
\end{equation}
according to numerical computations \citep{baumgarte1999b}.
This parameter is key for the outcome of the collapse,
since a value $<1$ allows for the whole mass to form a black hole without angular momentum loss.

Monolithic models have been used as initial conditions to simulate the final collapse of SMSs
\citep{saijo2002,saijo2004,liu2007,shibata2016b,sun2017,sun2018,li2018,uchida2017}.
A black hole is found to form, containing 90 -- 95\% of the stellar mass, and a typical spin parameter $a\sim0.7$.
The remaining gas is maintained outside the horizon by the centrifugal force and takes the geometry of a torus,
before relativistic jets are launched.
Interestingly, relaxing the constraint of solid rotation allows for more angular momentum to be contained in the star,
which results in more massive torus and more energetic jets \citep{new2001,sun2018}.

\qquad

\begin{figure}
\includegraphics[width=0.7\textwidth]{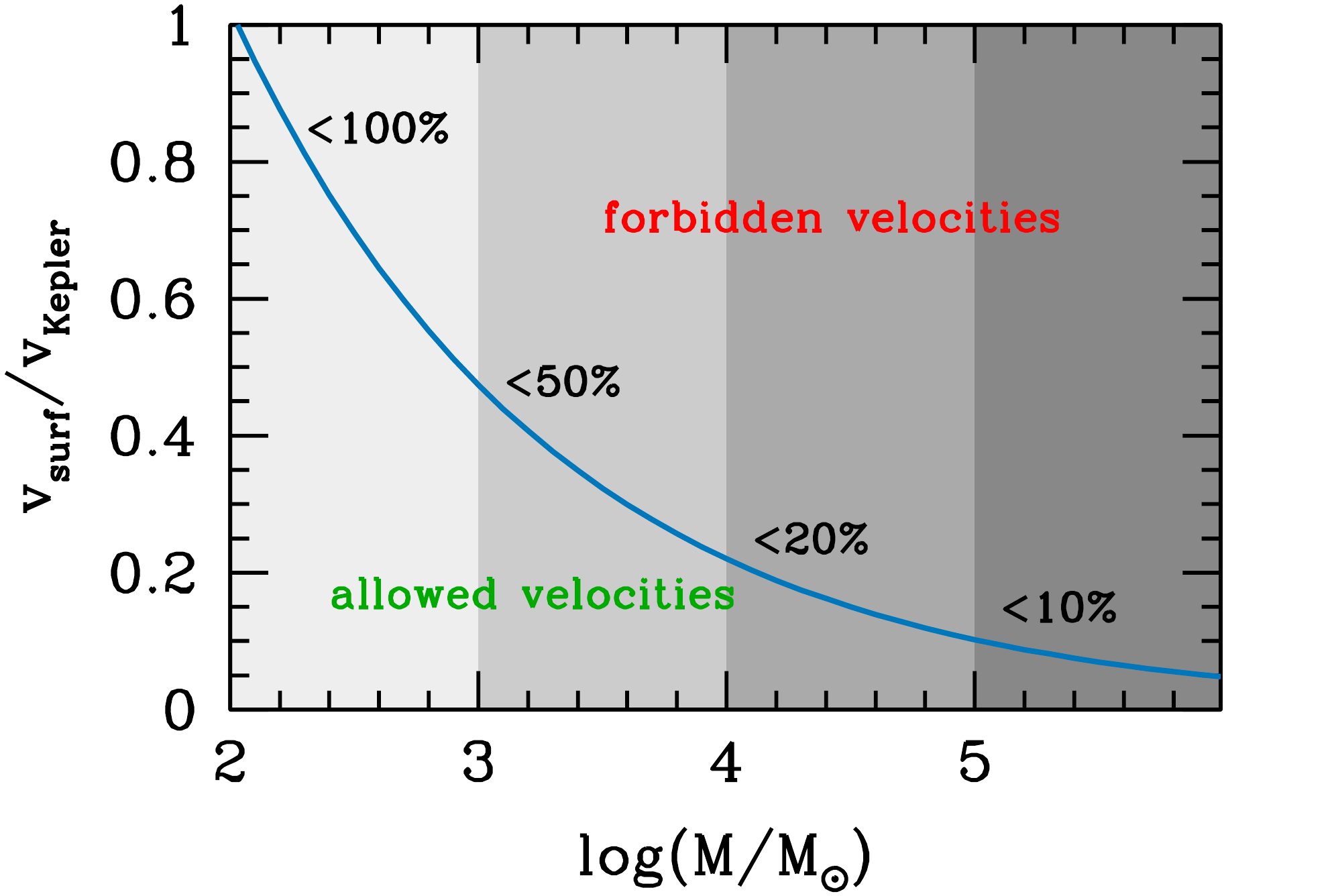}\label{fig-omgam}
\caption{Constraint from the $\Omega\Gamma$-limit on the surface rotation velocity of SMSs, from \cite{haemmerle2018b}.}
\end{figure}

When accretion at $\gtrsim0.1$ \Mpy\ is accounted for,
the rotational properties of SMSs are at the opposite extreme of those of monolithic models
\citep{haemmerle2018b,haemmerle2019a}.
First, since accretion stabilises the star against convection, solid-body rotation is not ensured,
and the highly non-homologous contraction of the stellar interior leads to strong differential rotation,
with a core rotating $10^4-10^5$ times faster than the surface in terms of frequency.
Second, building a supermassive hydrostatic object by accretion
requires its surface to rotate always at less than 10-20\% of the Keplerian velocity (Fig.~\ref{fig-omgam}).
Thus, accreting SMSs must be slow rotators.

This result is a consequence of the $\Omega\Gamma$-limit \citep{maeder2000},
i.e. the fact that the break-up velocity of massive stars is lowered by the effect of their strong radiation pressure,
compared to Keplerian velocity.
Break-up is reached when any contribution from gas pressure would destroy the star.
For low-mass stars, this condition translates into the equality between gravity and the centrifugal force,
which corresponds to the Keplerian limit:
\begin{equation}
{GM\over\RRec}={\vvcrit\over\Rec}		\qquad\Longleftrightarrow\qquad
\vcrit=\sqrt{GM\over\Rec}=\sqrt{{2\over3}{GM\over R}}=\vkep,
\end{equation}
where \Rec\ is the equatorial radius at break-up,
which according to the Roche model is 1.5 times the polar radius $R$, nearly unchanged by rotation.
The Roche model is always accurate for SMSs, thanks to the large density contrasts in their interiors.
For massive stars, one has to account for the effect of radiation pressure in the mechanical equilibrium,
and the condition for break-up becomes
\begin{equation}
{GM\over\RRec}={\vvog\over\Rec}-{1\over\rho}{\diff\prad\over\diff r},
\end{equation}
where \prad\ is the radiation pressure.
If we neglect the deformation of the star
and use the equation of radiative transfer $F=-{c\over\kappa\rho}{\diff\prad\over\diff r}$,
this condition gives a break-up velocity
\begin{equation}
\vog=\vcrit\sqrt{1-{\kappa L\over4\pi cGM}}=\vkep\,\sqrt{1-\Edd},
\end{equation}
where \Edd\ is the Eddington factor.
This expression shows that in the limit $\Edd\rightarrow1$ (Eddington limit), the corrected critical velocity goes to 0,
i.e. stars that are nearly Eddington must be slow rotators.
When accounting for the anisotropy of the flux, in the limit $\Edd\rightarrow1$, the expression becomes
\begin{equation}
\vog={3\over2}\,\vcrit\sqrt{1-\Edd}
\end{equation}
Interestingly, for $\Edd\lesssim0.6$, one has $\vog>\vcrit$.
This is a consequence of the von Zeipel theorem \citep{vonzeipel1924},
i.e. the reduction of the flux at the equator (where the centrifugal force is the strongest)
due to the decrease of effective gravity.
But for $\Edd\gtrsim0.6$, we see that $\vog<\vcrit=\vkep$,
i.e. the break-up velocity is lowered compared to the Keplerian limit.

Forming a supermassive hydrostatic object by accretion requires the constraint from the $\Omega\Gamma$-limit
to be satisfied all along the accretion phase, for the in-falling gas to be incorporated by the star.
As an extreme case of the $\Omega\Gamma$-limit,
accreting SMSs cannot rotate faster than 10 -- 20\% of their Keplerian velocity (Fig.~\ref{fig-omgam}).
It follows that their structure is not impacted by rotation.
In addition, rotational mixing (shear diffusion and meridional circulation)
remains negligible due to the short evolutionary timescales.

The constraint on the surface velocity translates into a constraint on the accreted angular momentum,
which cannot exceed significantly a percent of the Keplerian angular momentum.
The angular momentum barrier is thus particularly strong for SMSs.
Two mechanisms might allow for efficient angular momentum loss in the accretion process:
gravitational torques \citep{begelman2006,Wise2008,Hosokawa2016}
and magnetic fields \citep{pandey2019,haemmerle2019a,deng2019,deng2020}.
Supermassive accretion discs are highly gravitationally unstable,
and departures from axial symmetry are pronounced.
It results in strong gravitational torques that enhance accretion
by removing angular momentum from the in-falling gas \citep{Hosokawa2016}.
In addition, the turbulence driven by gravitational instability are found to generate strong magnetic fields,
in a much more efficient way than magneto-rotational instability \citep{deng2019,deng2020}.
These processes might be self-regulating,
since any cut in accretion due to the centrifugal barrier will result in the accumulation of mass in the disc,
which shall enhance fragmentation and turbulence.
In this picture, the star will adjust on the fastest rotation velocity compatible with the $\Omega\Gamma$-limit,
and the curve of Fig.~\ref{fig-omgam} would give the expected rotation velocity of SMSs as a function of the mass,
not only the upper limit.
Magnetic fields might play a role at all the stages of the SMBH formation process.
In the early phase of the collapse of the mini-halo, magnetic fields of 0.1 nG are strong enough to avoid break-up
\citep{pandey2019}.
Such fields are consistent with primordial fields whose flux was kept frozen during the formation of the halo.
During the main accretion phase onto the star, magnetic coupling between the stellar surface and its winds
could remove the 99\% of the Keplerian angular momentum
provided the magnetic field at the stellar surface is about 10 kG \citep{haemmerle2019a}.
Such fields are too large to be driven by a convective dynamo, but they are consistent with primordial origins
and correspond to the upper limit of detected magnetic fields on massive stars,
which are suspected to be fossil fields \citep{townsend2005,rivinius2010,grunhut2012a}.

\section{Formation of massive black hole seeds}
\label{sec:MBHS}

Here we cover the different flavours of direct collapse scenarios that have been proposed in the literature. The formation of massive BH seeds from metal-free
Population III stars is covered in Section 1. In this section we will also  briefly discuss the growth, by accretion, of black hole seeds formed from Pop III
stars and from direct collapse routes in light of the observational constraints currently posed by observations of bright QSOs at $z > 5$.

\begin{figure}
\centering
\psfig{figure=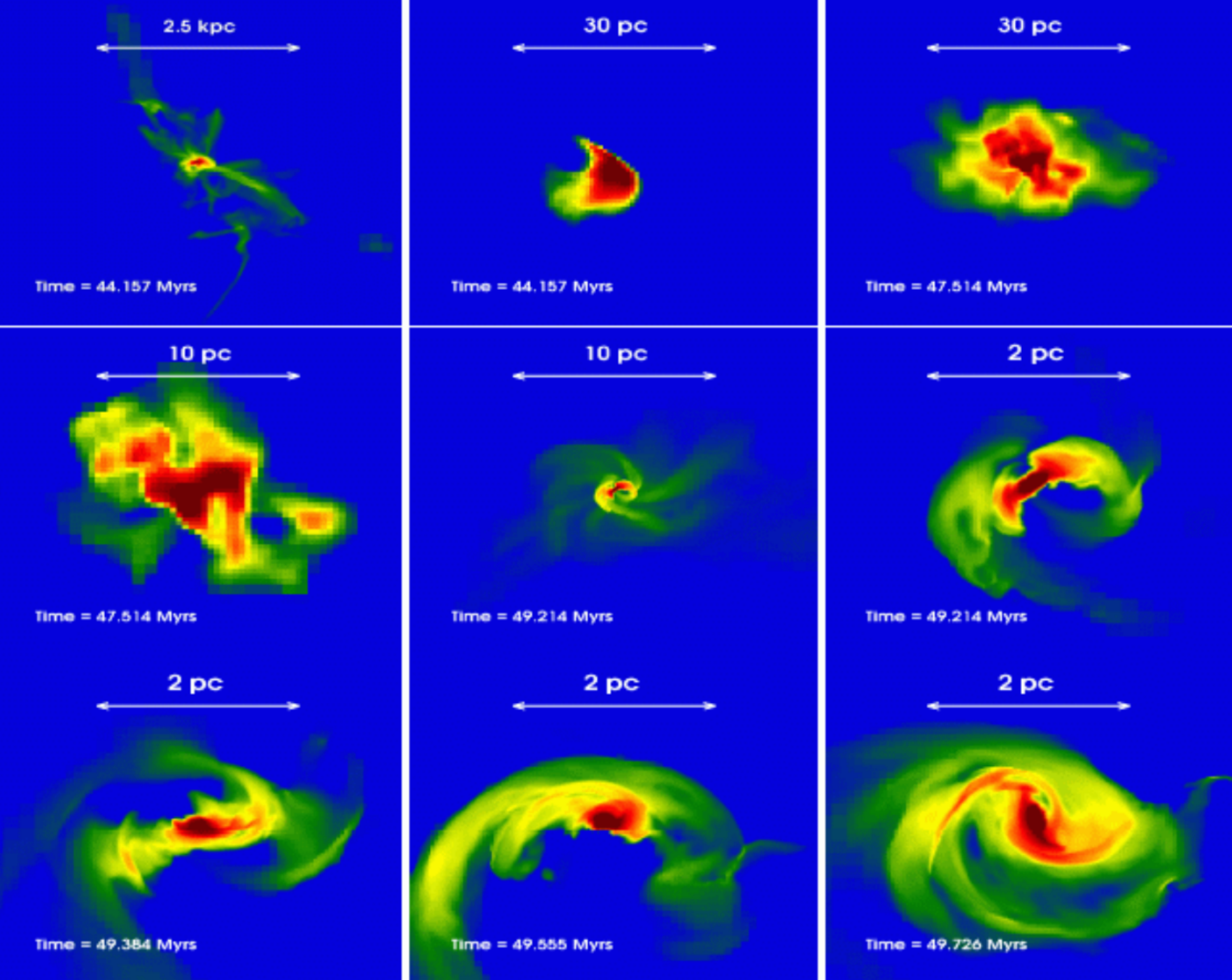,width=8.cm}
\vspace*{+0.0cm}
\caption{\small Color-coded density maps showing the evolution of the central
gaseous protogalactic disk inside a metal-free photodissociated atomic
cooling halo at $z \sim 203$ (adapted from \citealt{latif2013}). From top
to bottom increasingly smaller scales are shown, and the time evolution runs
from left to right. }
\label{figlatif}
\end{figure}

\subsection{The rapid rise of high-z QSOs and the challenge for light
black hole seed models}

The existence of extremely bright QSOs up to $z > 7$ \citep{fan06,Wu15,Banados18}
is in tension with BH formation scenarios in which BH
seeds start out light, of order a hundred solar masses as natural for the Pop III scenario (Sect.~\ref{sec:popiii}), and then grow gradually by Eddington-limited accretion. At $z > 7$, assuming the
standard LCDM cosmology, simply the lookback time is too short for a light BH seed to grow to $10^9$~\Ms, the BH mass inferred by assuming that the
source radiates at the Eddington limit (note that if accretion occurs at a fraction of Eddington, as suggested by other constraints for lower z QSOs, the
timescale problem is further exacerbated). 
We briefly recall here the arguments that support the need for a massive BH seed in order
to explain the rapid emergence of the bright high-z quasars.
The time $t(M_{\it BH})$ required for a black hole of initial mass $m_{seed}$ to reach a mass
$M_{\it BH}$, assuming Eddington-limited and continous accretion, is given by:
\begin{equation} \label{eqn:tBH}
 t(M_{\it BH}) = \frac{t_{\it Edd}}{f_{edd}} \frac{\epsilon_{r}}{(1-\eta)}
\ln \left( \frac{M_{\it BH}}{m_{seed}}\right),
\end{equation}
where $t_{\it Edd} \equiv c \sigma_{T}/(4 \pi G \mu m_{p})$ is the Eddinton time
($\sim 0.45$ Gyr for a pure hydrogen gas, with $\mu=1$),   $f_{edd}$ expresses
at which fraction of the Eddington limit the black hole is accreting, and
$\epsilon_{r}$ and $\eta$ are, respectively,
the radiative efficiency and accretion efficiency\footnote{The accretion
efficiency $\eta$ directly depends on the spin of the black hole, and reaches
the maximum value of $\eta \sim 0.42$ for maximally rotating Kerr black holes.
The radiative efficiency $\epsilon_r$  depends both on the type of accretion
and on the accretion efficiency: for radiatively efficient accretion events, one
can assume $\epsilon_r = \eta$.}. Assuming a radiatively-efficient accretion
event, we assume $\epsilon_r = \eta = 0.1$, a typical average value for Shakura-Sunyaev
thin disks \citep{Shakura73}.
Adopting a more realistic molecular weight per electron for a plasma at zero metallicity with
cosmic abundance of hydrogen (X = 0.75) and helium (Y = 0.25), $\mu =
1/(1  - Y/2)
= 1.14$, the Eddington time lowers to $t_{\it Edd} \sim 0.39$ Gyr. Note that the
precise value of the metallicity is  marginal in this calculation, as
metallicity has only  a
 minor effect on the value of the molecular weight (e.g., for a gas at solar metallicity,
$\mu=1.18$ instead of the value $\mu=1.14$ assumed here).
 The mass of the supermassive black holes powering the most luminous high-z
quasars is typically of the order of $M_{BH} = 10^9 M_{\odot}$, 
so that this is the value we assume
for the target mass. Note that this is a conservative 
choice as there is
at least one known  high-z QSO with black hole mass of $\sim 10^{10} M_{\odot}$, see \cite{Wu15}.
Further assuming that the seed black hole is able to
accrete continuously at the Eddington limit  ($f_{edd} = 1$), the timescale
necessary to reach the target mass is
$t_{(M_{BH})} = 0.4$ Gyr for
 a seed of $10^5 M_{\odot}$, while is   $t_{(M_{BH})} \sim 0.7$ Gyr for a black
holes  starting from $10^2 M_{\odot}$, the  average value for light
seeds produced by Pop III stars (see next paragraph and section 2).
In this latter case the growth timescale is
comparable to the age of the universe at $z = 7$ (assuming the Planck cosmology
of \citealt{ade2015}).

There are two alternative solutions, which are being extensively studied since a few years. In the first one,
the light BH seed is allowed to accrete above the Eddington limit, continuously or  episodically. 
Pop III remnants can form black hole seeds with masses up to $\sim1000$ \Ms\ (see Sect.~\ref{sec:popiii}).
The environments where Pop III BH seeds may form
at $z \sim  15-30$ have been extensively studied with numerical simulations, which
agree on the fact that such seeds grow too slow, at Sub-Eddington rates, as the gas in their
host mini-halo is efficiently photoionized, and eventually  expelled, owing to radiative feedback from the massive stellar progenitor
(eg., \citealt{Johnson2007,Wise2008}).  In particular, primordial stars in the mass range $140-260
M_{\odot}$ explode as pair-instability supernovae \citep{Heger2003},
liberating enough energy to effectively empty the mini-halos
of gas. Lower-mass Pop III stars would still damp energy, having a significant
impact on the ambient medium if they form in clusters, as it is likely \citep{Greif15}, owing to the shallow potential wells of mini-halos.
X-ray radiative feedback also suppresses further fueling of  gas, leading effectively to accretion rates as small as $\sim 1\%$ of the
Eddington rate, and thus stifling the growth of the BH seed well below the expectations from  the Eddington-limited accretion
model  considered above \citep{alvarez2009,Jeon12}.

Only later, when star formation has led
to metal enrichment, increasing the cooling rate, and the original host halo has grown significantly in mass to allow atomic cooling via Ly-$\alpha$ radiation, accretion can in principle restart (see also the recent review by \citealt{inayoshi2019}).

Super-Eddington accretion onto light seeds could then be achieved, in principle, in radiatively inefficient accretion disks or in accreting envelopes emitting highly anisotropically (see \citealt{mayer2019} for a review, and \citealt{inayoshi2019,Takeo18}).
Several studies, both at the level of radiation hydrodynamics and MHD simulations
of various accretion flow configurations (e.g. \citealt{Sadowski13}), and at the level of larger scale models at nuclear (proto) galactic scales with parametrized analytical accretion models,
such as the SLIM disk model, have been developed in the last decades. Alternatively,
the formation of a much heavier BH seed, with mass $> 10^4 M_{\odot}$ can be invoked, which could then grow in the required time
to billion solar masses in an Eddington-limited regime. This second route is what we will discuss thoroughly in the remainder
of this section, and goes by the name
of {\it direct collapse}. Direct collapse is a general scenario rather than a specific model. There
are indeed various models  proposed to generate a massive BH seed without starting from a conventional stellar progenitor, as recently reviewed in detail by \cite{woods2019}. Two of the models examined in the literature are covered in the next subsection.

\subsection{Direct collapse scenarios for the origin of massive BH seeds}

In direct collapse scenarios the initial BH  is not
resulting from the collapse of an ordinary star. Note
that, at high redshift ($z>20$) , for ordinary stars we mean Population III stars as described in the first section of this review. Population III stars would
produce a {\it light} BH seed with mass in the range
from a few tens to of order thousand solar masses,
according to the mass distribution of such early
population of metal-free stars.
In direct collapse scenarios the BH seeds that
arise have a mass of at least $10^4$ \Ms, but
can be even as large as $10^8$ \Ms\ in 
the most extreme models including rotation (Sect.~\ref{sec:SMSom})
or through {\it dark collapse} \citep{mayer2019}.
The formation of an SMS is usually postulated for the precursor stage, which then collapses into a massive BH seed,
by the GR radial instability, or because the SMS progenitor can never achieve a global hydrostatic equilibrium
(see Sect.~\ref{sec:SMSmax}).

\begin{figure}
\centering
\psfig{figure=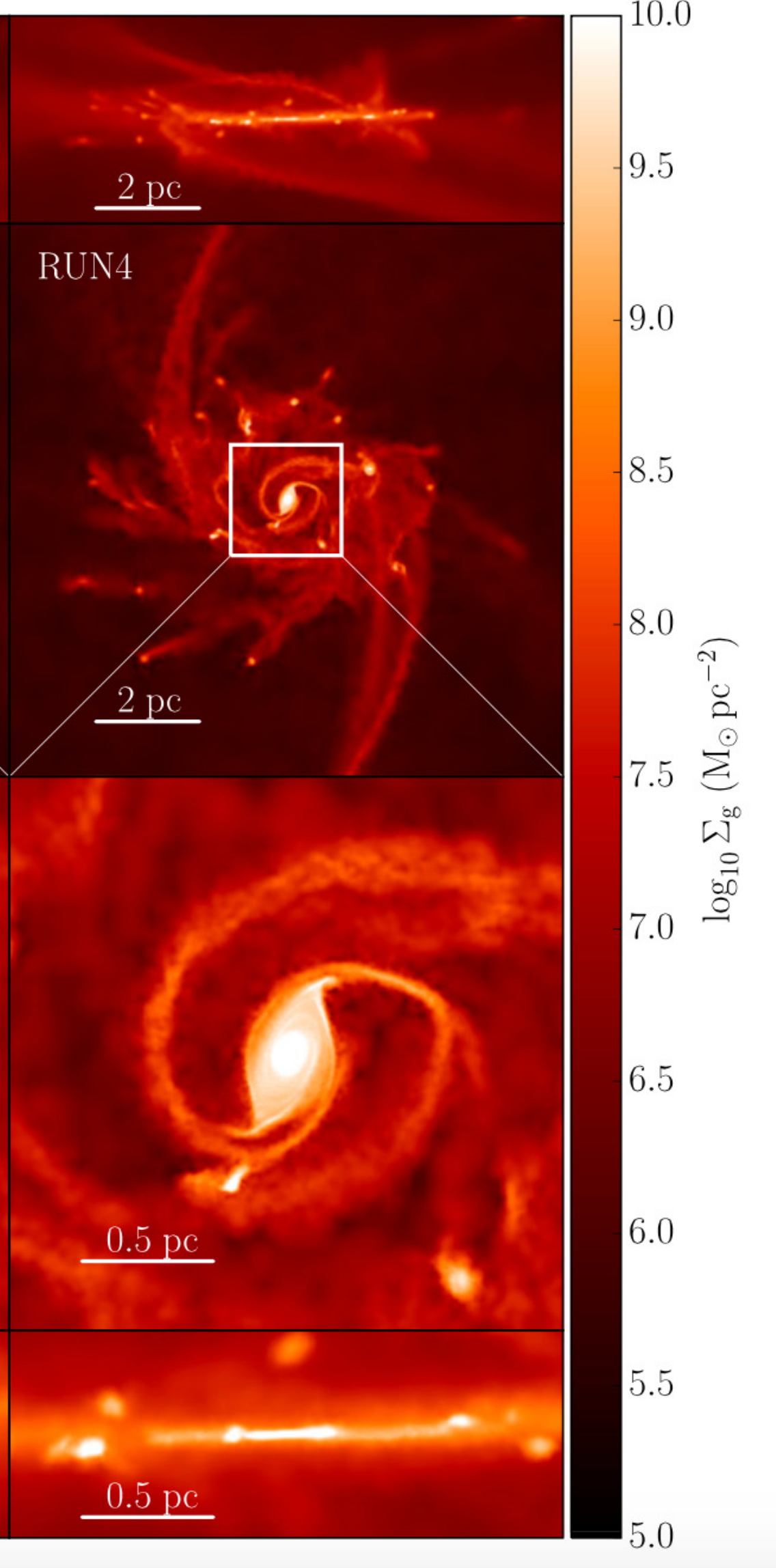,width=5.cm}
\vspace*{+0.0cm}
\caption{\small 
Face-on and edge-on projected gas density maps of the nuclear region of the merger at $t_0 + 5$ kyr ($t_0$ is the time
corresponding to the merger of the two central cores.)
showing a disk-like object with
radius 5 pc (first and second panel are edge-on and face-on, respectively), and a compact inner disk-like core less than a parsec in size (third
and fourth panels are face-on and edge-on, respectively). We show one particular run from the Mayer et al. 2015 paper, RUN4, which employed a modern
version of the SPH hydro solver using pressure-energy formulation (GDSPH), thermal and metal diffusion, and a Wedland kernel.
The figure is adapted from \cite{mayer2019}.
}
\label{fig1}
\end{figure}

\subsubsection{A crucial aspect: angular momentum transport in protogalaxies}

The first stage in the direct collapse scenarios involves gas skipping conventional star formation and reaching, instead,  extremely high densities, collecting a gravitationally
bound mass large enough to generate {\it a supermassive precursor object}. Note that we will assume this all happens in the galactic nucleus since QSOs are associated
to galactic nuclei, but most of our considerations are not necessarily restricted to the nuclear region only. 
In order to generate a supermassive precursor, irrespective on whether or not the object will remain in a protostellar-like regime before collapsing, or will shine,
even briefly,  as an SMS, enormous quantities of interstellar gas have to be delivered to the galactic nucleus from galactic scales. This gas then has to become
gravitationally bound and produce the precursor. These steps, necessary for the precursor stage, are a crucial aspect of direct collapse models and, while various
propositions have been made, the jury is still on what is the actual mechanism that prevails, if any.

If there is general consensus on one aspect that is on the broad scenario for how
angular momentum can be lost in protogalaxies. This is believed to happen primarily
via gravitational torques in primordial gas-rich galaxies (e.g. \citealt{begelman2006,lodato2006}, perhaps boosted by violent phenomena such as galaxy
mergers \citep{mayer2010,mayer2015}. In order for such torques to be effective at removing angular momentum from 
gas the galactic disks, and in particular their nuclear regions, need to be in
a marginally self-gravitating state. This corresponds to a regime in which the
Toomre Q parameter \citep{toomre1964}, which measures the susceptibility of the disk to become
unstable to its own self-gravity, must hover just above unity. In an entirely
gaseous disk the Toomre parameter is defined as $Q = c_s \kappa/\pi G \Sigma$,
where $c_s$ is the sound speed, $\kappa$ is the epicyclic frequency of gas motion,
$\Sigma$ is the gas surface mass density, and $G$ is the gravitational constant.
The definition is strictly local, although it is customary to use it also globally
or expressing it using average properties in the disk. If Q drops below unity in
most of the disk, gas will fragment copiously, producing substructure (clouds, and then stars
from clouds possibly), thus interrupting the large scale global torquing action
of spiral density waves by breaking their coherence. If, by heating or mass loss, it increases comfortably above unity
($Q > 2$) the disk will become gravitationally stable, and thus not capable anymore
to transfer angular momentum and re-shuffle mass.
If $Q$ remains low enough to allow self-gravity to excite non-axisymmetric structures, such as bars
and spiral arms, efficient angular momentum transport can occur as gravitational torques by such
structures behave as an effective viscous drag acting on the gas flow.
Numerical simulations of protogalaxies have indeed shown that gas inflow rates are triggered 
by non-axisymmetric instabilities, such as spiral arms \citep{latif2013} and bars-in-bars instabilities 
\citep{shlosman2011,choi2013}. A clear example is shown in Figure 8.

Neglecting gas accretion, the stability of a 
gas-dominated protogalactic disk will be determined by the two conditions valid for any self-gravitating gas disk. Namely, its surface
density must be  low enough to maintain the 
Toomre Q above unity,
and the cooling time has to be longer
than the local orbital time \citep{Gammie01}.
For the latter condition, to so-called
{\it Gammie criterion)},
simulations
using improved numerical resolution have
demonstrated convergence as a function of resolution after a decade of discordant results
\citep{deng2017}.
If the cooling time
drops below the local orbital time the Toomre
Q will decrease and eventually drop below unity,
entering the fragmentation regime.

Models of direct collapse have thus focused
on ways to avoid rapid cooling of the protogalactic disk in order to maintain sustained gas
inflows as opposed to fragmentation. At sufficiently high redshift ($z > 15-20$) gas
in typical dark matter halos was still  metal-free (with the exception of rare high
density peaks), and radiative cooling was occurring either
via molecular lines, predominantly via roto-vibrational transitions of the $H_2$ molecule,
or, in sufficiently massive halos ($M_{vir} >10^7 M_{\odot}$), via atomic hydrogen line
cooling. If $H_2$ cooling is suppressed, the
protogalactic disk would not cool below a temperature
of $\sim 10^4$ K, thus evolving isothermally
and remaining stable to fragmentation,
as found in the  semi-analytical model
of \cite{lodato2006}.   Achieving suppression
of molecular cooling has thus become the main objective in direct collapse models.

\begin{figure}
\centering
\psfig{figure=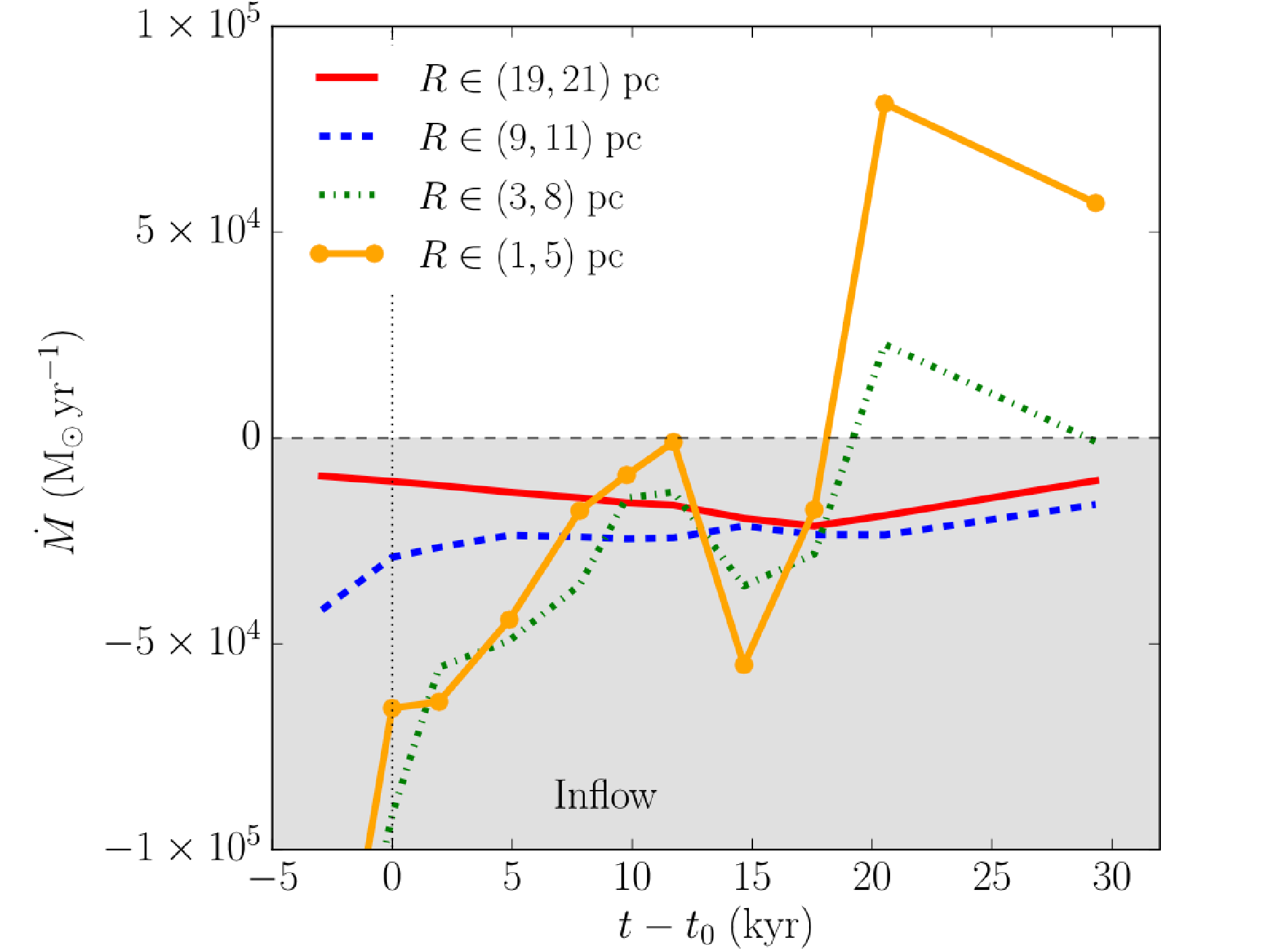,width=9.cm}
\vspace*{+0.0cm}
\caption{\small
Time evolution of the gas accretion rate at different radii from the center of the merger remnant.
The accretion is computed inside
cylindrical shells of inner and outer radii marked in the legend and vertical thickness 2 pc. RUN4 from \cite{mayer2015} is used, which is also
shown in Figure 3. Adapted from \cite{mayer2015}.)
}
\label{fig1}
\end{figure}

\subsubsection{Routes to direct collapse: metal-free atomic cooling halos and mergers of most massive metal-enriched primordial galaxies}

In the last couple of years the community has made progress in studying both the conventional route to direct collapse,
which involves gas inflows in metal-free atomic cooling halos (e.g. \citealt{latif2013,Regan09,Regan17}) and
other scenarios, such as the merger-driven model \citep{mayer2010,bonoli2014,mayer2019}, or models whereby colliding streams in proto-halos at very high-z \citep{hirano2017}
or streaming velocities \citep{Schauer17} play a role in providing favourable conditions.

The most conventional route to suppress $H_2$ cooling has been
to invoke nearby sources of Lyman-Werner photons, which can efficiently dissociate
$H_2$. This could happen by invoking a {\it proximity} condition, as a sufficiently nearby star forming (proto)galaxy
could be responsible for a high dissociating flux. The idea was proposed originally by \cite{Dijkstra08},
and explored first assuming a constant local dissociating Lyman-Werner background, to find out the required
level in order to accomplish  $H_2$ dissociation (e.g. \citealt{latif2013}), and, lately, computing self-consistently
the spatially and temporally dependent dissociating flux from sources in galaxies forming within hydrodynamical
cosmological simulations of the early Universe before the reionization epoch \citep{Regan17,regan2019,wise2019}.
A large body  of theoretical and numerical work has led to the notion that the required Lyman-Werbner flux is roughly
1000 times higher than the reference intensity value $J_{21} =  1 × 10^{-21}$ erg $s^{-1}$ cm$^{-2}$ Hz$^{-1}$ sr$^{-1}$
(see review by \citealt{latif2019}), orders of magnitude above the expected mean background flux value (e.g. \citealt{ahn09}).

\begin{figure}
\centering
\psfig{figure=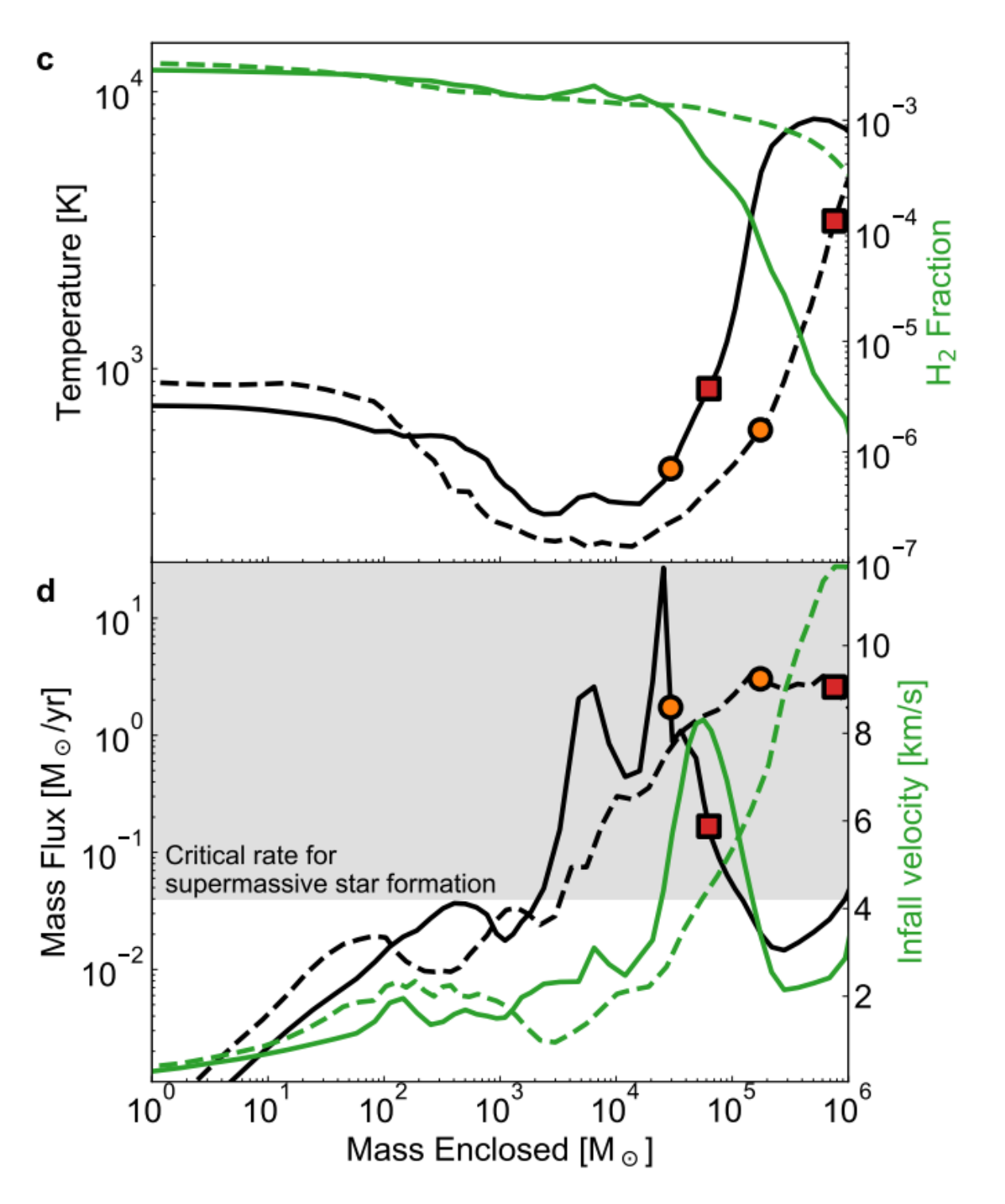,width=6.cm}
\vspace*{+0.0cm}
\caption{\small The figure, adapted from \cite{wise2019}, shows, for their high resolution re-simulations of two rare peaks in the {\it Renaissance}
cosmological simulation (indicated respectively, by the dashed and solid lines); gas temperature (c, black lines), $H_2$
fraction (c, green lines), radial mass infall rates (d, black lines), and radial infall velocities (d, green lines) . The orange circles
indicate the Jeans mass and associated quantities at that mass scale when the object first becomes gravitationally unstable, whereas the red squares represent the same
quantities at the end of the simulation.}
\label{figwise}
\end{figure}

Current state-of-the-art simulations, such as {\it Renaissance} (\cite{wise2019}, see Fig.~\ref{figwise}), showed that the mechanism can be  successful in very rare objects
forming relatively late, at $z < 20$, as it requires that, in addition to the proximity requirement, 
on one hand, sufficient metal-pollution to power gas cooling and intense
star formation  in the neighboring galaxy, and, on the other end, that the target system is still metal-free by that time.
In the {\it Renaissance} cosmological volume, nearly 40 Mpc on a side, out of which subvolumes encompassing high density
regions were re-sampled at higher resolution, a few tens of candidate systems that remain star-free and metal-free
by the onset of the atomic cooling halo phase were found \citep{regan2019}.
Most importantly, the {\it Renaissance} runs also showed that dynamical heating by mergers and
accretion is contributing significantly to maintain a warm nearly isothermal core in the protogalaxy, hence relaxing the requirement
on the critical Lyman-Werner flux. The few tens of atomic cooling halos that remain metal-free and star-free by the end of the
simulation have indeed been exposed to a mean Lyman-Werner flux just comparable to $J_{21}$, yet the simulations are not conclusive as 
they lack the resolution to exclude fragmentation of the protogalactic cores into clusters of Pop III stars \citep{regan2019}.
In addition, even if fragmentation would  not be an issue, the formation of a direct collapse BH seed would require 
sustained nuclear accretion rates at $> 0.1$ \Mpy\ in the protogalactic nucleus to produce an SMS (see section 2),
or even a quasi-star \citep{begelman2010}, for timescales $> 10^6$ yr (namely at least longer than the expected lifetime
of the SMS or quasi-star). In at least two of the highest resolution re-simulations of selected
rare high density peaks in the {\it Renaissance} volume, such high inflow rate 
conditions seem indeed to be fulfilled towards the end of the simulation, almost 1 Myr
after the beginning of gas collapse in the halo \citep{wise2019}. 
However, gas accumulation occurs on separate gravitationally bound clumps that, by the
same time, achieve a mass only an order of magnitude higher than that of Pop III stars,
between $1000$ and $10^4$ \Ms. With such the latter mass range for the precursors  the resulting massive BH seed would  still be
too light to explain the most massive of the high-z QSOs ($M_{BH} \sim 10^{10} M_{\odot}$)
by Eddington limited accretion.

Mergers can also temporarily 
enhance gas inflow rates by removing angular momentum via strong tidal torques and gas shocking, these  being considerably more intense
than internal torques due to disk instabilities.
The effect of dynamical heating brings us to the other proposed mechanism, namely heating by gas accretion and shocking induced
mergers  of much more massive and more mature galaxies at somewhat later epochs, $z \sim 8-10$.
At this stage the
first massive galaxies residing in halos with
$M_{vir} \sim 10^{12}$ \Ms\ do appear \citep{mayer2010}. This mass scale is interesting since it is consistent with the
virial mass expected for the hosts of the high-z QSOs given their clustering amplitude and low abundance \citep{Volonteri2008}.

In this second scenario, since it occurs much later, gas is already metal-enriched, yet simulations have shown that cooling is still
inefficient as compressional heating and shocks in equal mass mergers are sufficient to counteract cooling by maintaining the
temperature of the pc-scale disky core emerging in the merger remnant at 5000 -- 8000 K, allowing only sporadic fragmentation at its outskirts (\citealt{mayer2015}, see
Figure 10).
Most importantly, in this case radial gas inflows, and thus gas accretion rates onto the core, occur at the phenomenal rate
in the range $1000 - 10^4$~\Mpy\ for nearly $10^5$ yr after the merger (Figure 11), allowing the formation of an ultra-compact supermassive
rotating gas core that thus  meets naturally the conditions to become an SMS (see Sect.~\ref{sec:SMSmax}).  This allows the central core to reach a mass of $\sim
10^9 M_{\odot}$ in less than $10^5$ yr. The central core is fast rotating and gravitationally
bound, and with such a high mass it can form, in principle, a supermassive BH seed which
would have no problem in reaching the mass of the brightest QSOs in less than hundred million
years even with accretion sustained at a fraction of Eddington \citep{bonoli2014,mayer2019}.

The reason why inflow rates are so much higher in the merger-driven collapse is partially because major mergers produce the strongest tidal torques
and shocks, as they give rise to the largest displacement of mass, and partially because the halos involved in this scenario
are, by construction, much more massive than atomic cooling halos. Indeed, to zeroth order, neglecting residual angular momentum in the flow, the
radial gas flow will displace mass with amplitude $\dot M \sim V_{vir}^3/G \sim M_{vir}$ (where $M_{vir}$ and $V_{vir}$ are the
virial mass and virial velocity of the host halo, respectively), with $M_{vir}$ increasing by more than
a factor of 1000 between atomic cooling halos ($M_{vir} \sim 10^{9-10} M_{\odot}$) and high-sigma peaks at $z \sim 8-10$ possessing
$M_{vir} \sim 10^{12} M_{\odot}$ (see \citealt{bonoli2014,mayer2019} for details).
New multi-scale
merger simulations of high density peaks reaching parsec resolution with fully cosmological initial conditions (from the previously
published MassiveBlack runs of \citealt{Feng11}) appear to confirm the existence of such prominent gas inflows (Capelo, Mayer
et al., in preparation). With such extreme conditions it is also possible that the SMS stage is skipped, and rather a BH seed
nearly as massive as the entire supermassive gas core could form either because it becomes compact enough to collapse due to the
General Relativistic  radial instability \citep{mayer2019}, or because it can never find an hydrostatic equilibrium
configuration (see Sect.~\ref{sec:SMSmax}), a route refereed to as {\it dark collapse}.
In this latter case an ultra-massive BH seed would arise, as
massive as nearly the entire precursor core of $\sim 10^9 M_{\odot}$, since
numerical relativity simulations of rotating polytropic SMS/clouds reaching a critical compactness show that the radial GR instability 
triggers a global collapse of about half of the total mass into a central black hole (e.g. \citealt{shibata2002}), or even
into a binary system of two black holes if the massive precursor  becomes bar unstable and fragments \citep{reisswig2013}.

Observationally, both the conventional direct collapse scenario based on $H_2$ dissociation
in metal-free atomic cooling halos and the merger-driven scenario gives rise to a number
of direct collapse BHs that, while low in absolute sense, it is still high enough to accomodate
the abundance of high-z QSOs \citep{mayer2019,latif2019}. However, a sensible
quantitative comparison with observations will have to take into account the detailed mechanism 
of BH seed formation from the precursor, eg whether an SMS, a quasi-star, or neither of the two, will form prior to collapse (see \citealt{woods2019}), and the duty-cycle of the resulting AGN.

\section{Reionization after {\it Planck} and before {\it JWST}}
\label{sec:reionization}

The  transformation of cold neutral intergalactic hydrogen into a highly ionized warm plasma marks the end of the dark ages and the beginning of the age of galaxies. In popular cosmological scenarios, massive stars within early protogalaxies, aided perhaps by a population of accreting seed black holes, generated the UV and X-ray radiation that reheated and reionized the Universe. Observations of resonant absorption in the spectra of distant quasars show that hydrogen reionization was still ongoing at $z\sim 6$ and  fully completed by redshift 5.5 \citep{fan06,becker15}. An early onset of reionization appears to be disfavored by the final, full-mission {\it Planck} CMB anisotropy analysis, which yields an integrated Thomson  scattering optical depth of $\tau_{\rm es}=0.054\pm 0.007$ \citep{planck18}. And while the evolution of the ionization and thermal state of the $z\gta 6$ Universe is being constrained by a number of direct probes, from the damping wing absorption profiles in the spectra of quasars \citep[e.g.,][]{banados17}, through the luminosity function (LF) and clustering properties of \Lya\ emitting galaxies \citep{ouchi10,schenker14}, to the flux power spectrum of the \Lya\ forest \citep{onorbe_ps17}, many pivotal aspects of this process remain highly uncertain.

\begin{figure}
\centering
\psfig{figure=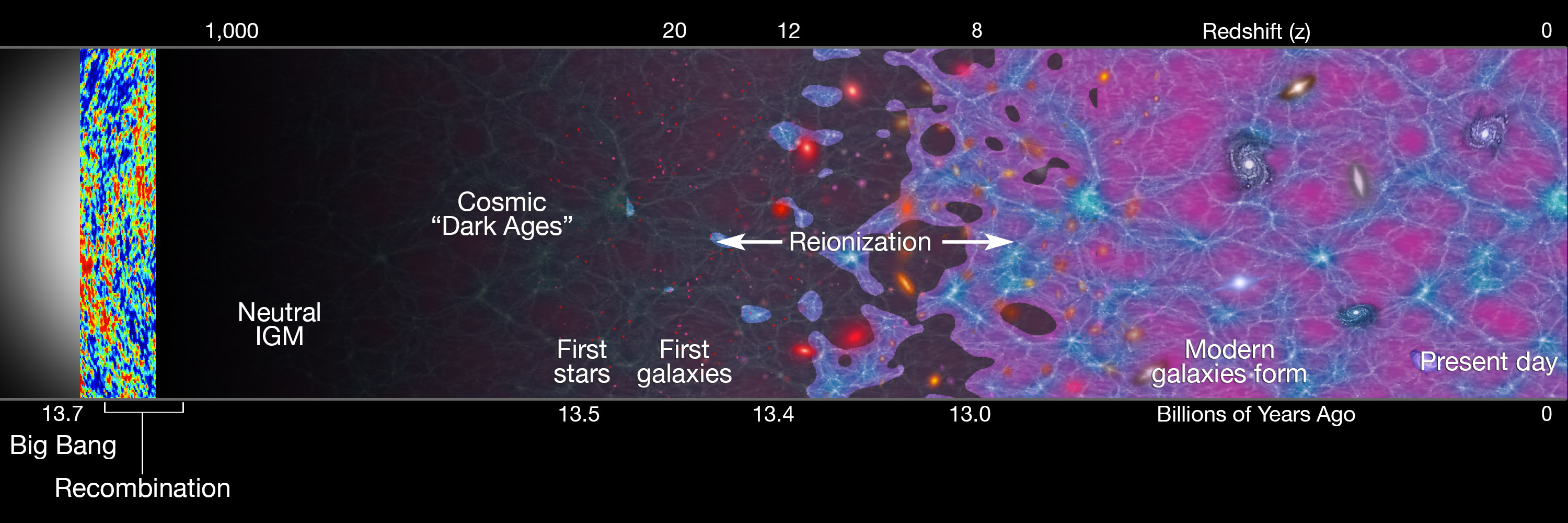,width=12.cm}
\vspace*{+0.0cm}
\caption{\small Cosmic Reionization Fundamentals. After recombination at $z\sim 1100$, hydrogen in the IGM remained neutral until 
the first stars and galaxies formed at $z\sim 15-20$. These primordial systems released energetic UV photons capable of ionizing 
local bubbles of hydrogen gas. As the abundance of these early galaxies increased, \HII\ bubbles overlapped and progressively larger 
volumes became ionized. This reionization process completed at $z\sim 6$, approximately 1 Gyr after the Big Bang. At lower redshifts, 
the IGM remains highly ionized through radiation provided by star-forming galaxies and the gas accretion onto supermassive black holes 
that powers quasars. (From \citealt{robertson10}.)
}
\label{fig1}
\end{figure}

Over the last few years, {\it HST} deep optical and NIR imaging data have led to the discovery and study of some of the earliest-known galaxies at redshifts $z=7-10$, corresponding to a period when the Universe was only $\gta 500$ Myr old \citep[e.g.,][]{finkelstein15,livermore17,oesch17}.
The faint-end slope of the rest-frame UV LF of these high-redshift sources has been measured to be very steep ($\alpha\approx-2$), such that the integrated UV luminosity density is contributed equally by each decade in galaxy luminosity, and therefore may depend on whatever physics regulates star formation in the faintest dwarfs \citep[see, e.g.,][and references therein]{bouwens12,atek15,ishigaki17}. 
The importance of star-forming galaxies for reionization further depends on the Lyman-continuum (LyC) photon yield per unit UV luminosity 
and on the fraction of these photons that can escape into the IGM \citep{madau99,robertson10}. 
Galaxy-dominated scenarios that satisfy many of the observational constraints can be constructed if one extrapolates the UV LF 
to sufficiently faint magnitudes ($M_{\rm UV} \sim -13$) assuming a time- and luminosity-averaged escape fraction of $\sim$ 15\% at $z\gta 6$ \citep[e.g.,][]{haardt12,robertson15,bouwens15,madau17}. 
The presence of moderate- and low-luminosity AGNs at early times can change inferences on the role of galaxies in reionization 
\citep[e.g.,][]{madau15,giallongo19}, and the differing effective temperature of the radiation between stars and AGN can impact 
the thermal history of the reionized IGM \citep{daloisio17}. At low metallicity, stellar populations in early galaxies may be 
characterized by an increased LyC photon production from binary stars compared to the present-day Universe \citep{stanway16}. 
The enhanced LyC photon yield may boost the equivalent width of nebular emission lines ([O III], H$\alpha$, H$\beta$) 
\citep[e.g.,][]{smit14}, which are redshifted into the rest-frame wavelength range relevant for the James Webb Space Telescope ({\it JWST}) NIRSpec instrument.
The impact of nebular emission on the spectra of galaxies is also regulated by the escape fraction, and forthcoming measurements of 
the rest-frame UV continuum slope with the equivalent width of the H$\beta$ emission line may allow the 
identification of galaxies with extreme LyC leakage up to redshift 9 \citep{zackrisson17}. 

\begin{figure}[t!]
  \minipage{0.34\textwidth}
  \raisebox{0.33cm}{\centering{\includegraphics[trim=0 0 0 0,clip,width=1.00\textwidth]{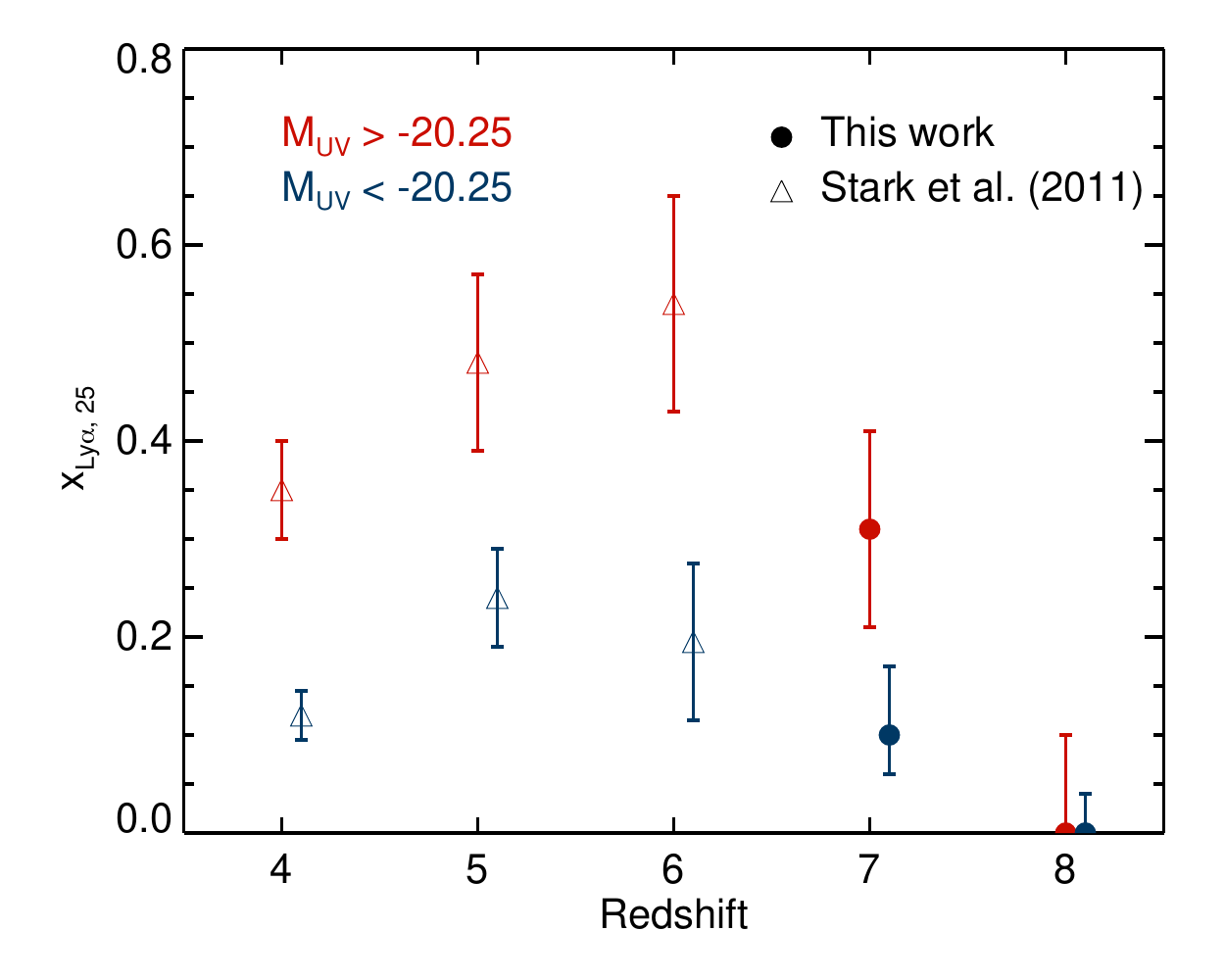}}}
  \endminipage
  \hskip -0.2cm
  \minipage{0.33\textwidth}
  \vskip -0.2cm
  \raisebox{0.0cm}{\includegraphics[trim=5 0 10 15,clip,width=1.05\textwidth]{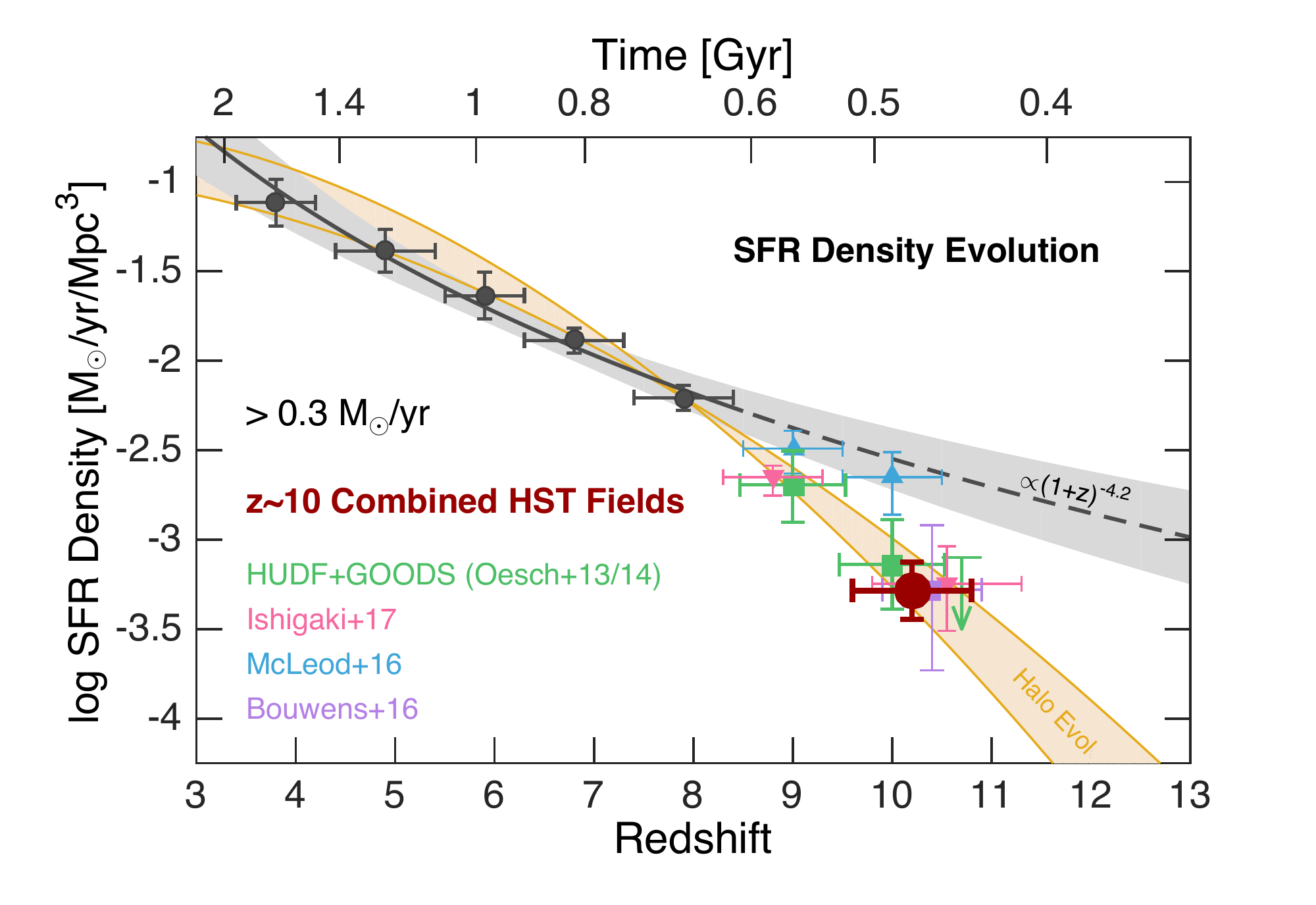}}
  \endminipage
  \hskip 0.1cm
  \minipage{0.33\textwidth}
  \raisebox{0.0cm}{\centering{\includegraphics[trim=7 0 10 20,clip,width=1.03\textwidth]{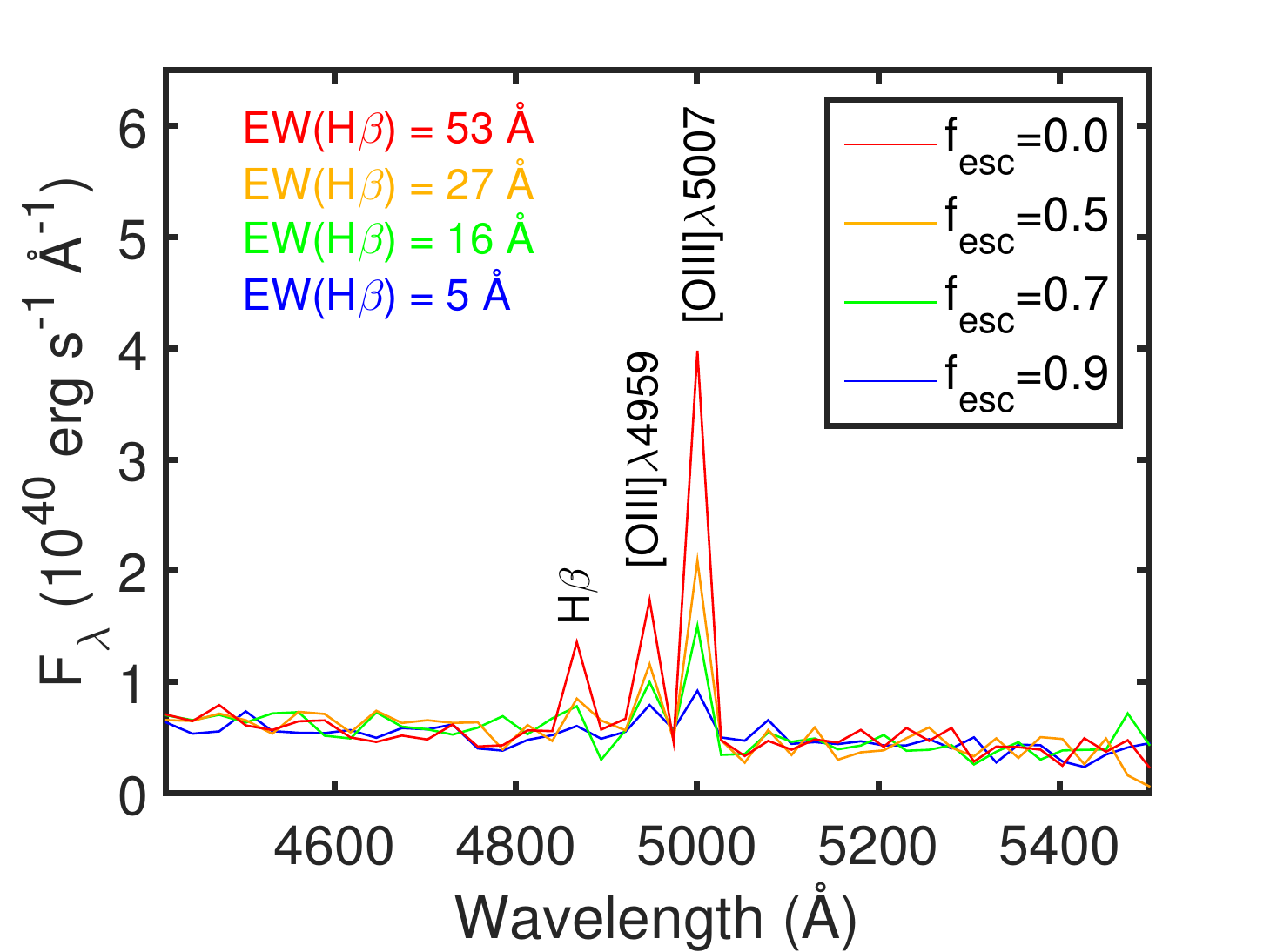}}}
  \endminipage
  \hfill
  \vspace{-0.2cm}
 \caption{\small Observational Probes of Reionization: I.
\textit{Left:} Fraction of Lyman break galaxies that display \Lya\ in emission at an EW 25\AA\ plotted as a function
of redshift. The values at $z = 7$ and 8 reflect differential measurements with the data at $z = 6$. (From \citep{schenker14}.)
\textit{Center:} Evolution of the cosmic star formation rate (SFR) density at high redshift integrated down to a UV luminosity of $M_{\rm UV} = -17.0$
(SFR $\gta 0.3 M_\odot \,{\rm yr}^{-1}$). New measurements from the combination of all {\it HST} fields (filled dark red circle) suggest a  
rapid, accelerated evolution of this quantity between redshifts 8 and 10. The orange shaded region shows the relative evolution 
of the cumulative DM halo mass function integrated down to $\log{M_{\rm halo}\slash M_{\odot}} = 9.5-10.5$. (From \citep{oesch17}.)
\textit{Right:} Synthetic spectrum of a single $M_\ast \approx 7 \times 10^8 M_\odot$, $m_{\rm AB}=27$, dust-free star-forming 
galaxy for different values of the LyC escape fraction $f_{\rm esc}$: 0.0 (red), 0.5 (orange), 0.7 (green) and 0.9 (blue).  The spectrum 
has been degraded to the spectral resolution of the {\it JWST}/NIRSpec $R = 100$ prism, and with noise level corresponding to a 10h exposure. 
The $f_{\rm esc}$ parameter mainly affects the strength of the emission lines relative to the continuum, and this effect remains detectable 
for the strongest nebular lines.  (From \citep{zackrisson17}.)
}
\label{figEOR1}
\end{figure}

\begin{figure}[t!] 
  \minipage{0.32\textwidth}
  \raisebox{0.0cm}{\centering{\includegraphics[trim=0 0 0 10,clip,width=\textwidth]{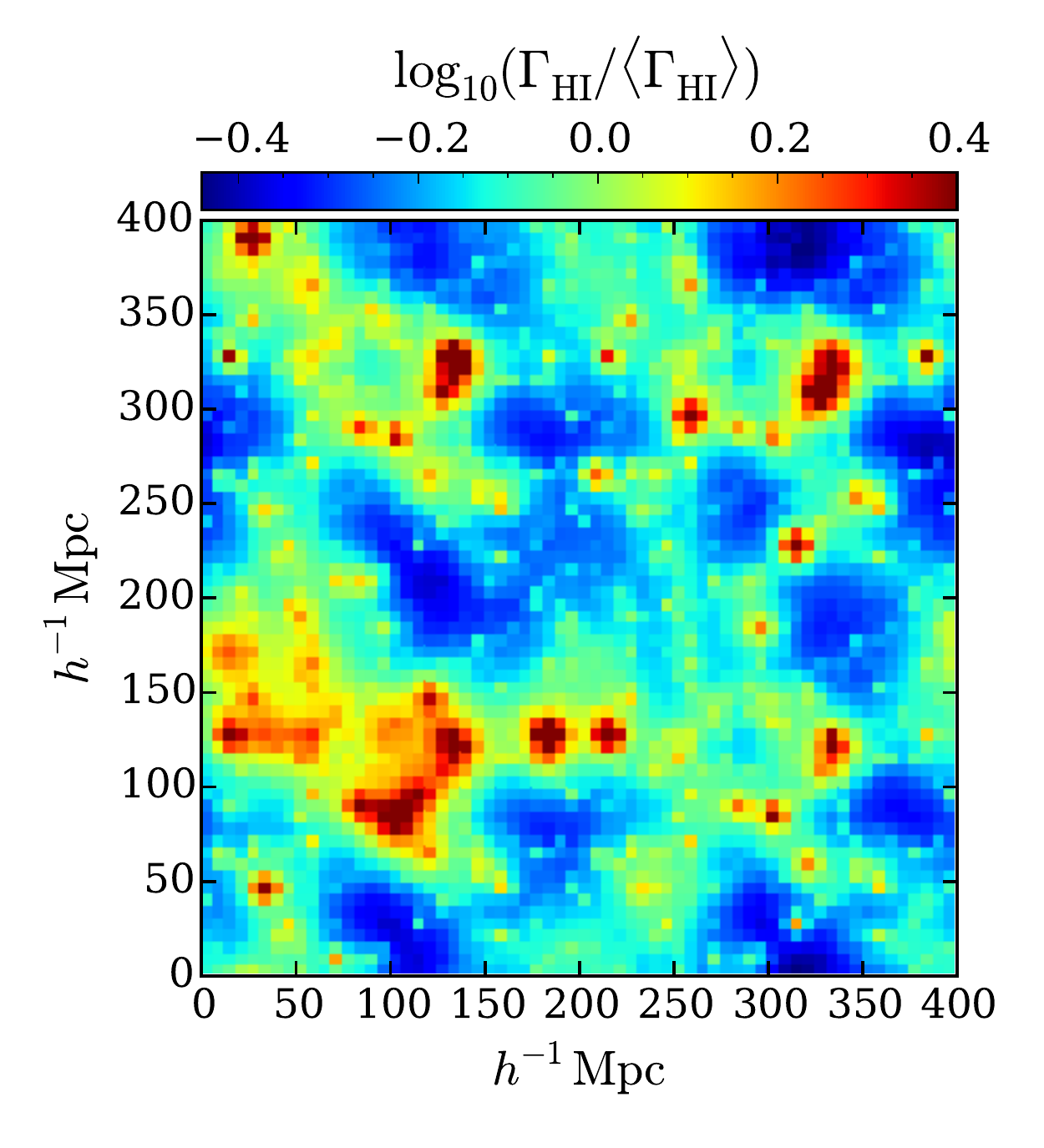}}}
  \endminipage
  \minipage{0.32\textwidth}
  \hskip -0.2cm
  \raisebox{0.0cm}{\centering{\includegraphics[trim=0 0 0 0,clip,width=\textwidth]{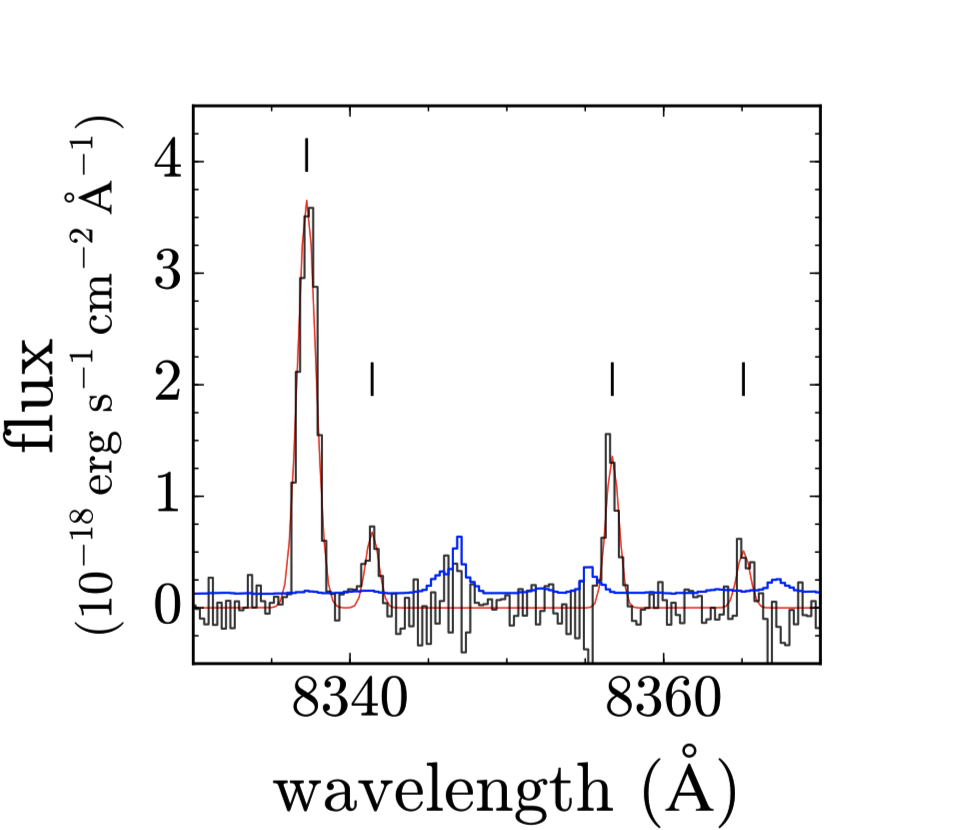}}} 
  \endminipage
  \minipage{0.32\textwidth}
  \raisebox{0.0cm}{\centering{\includegraphics[trim=0 0 0 0,clip,width=0.95\textwidth,height=4.0cm]{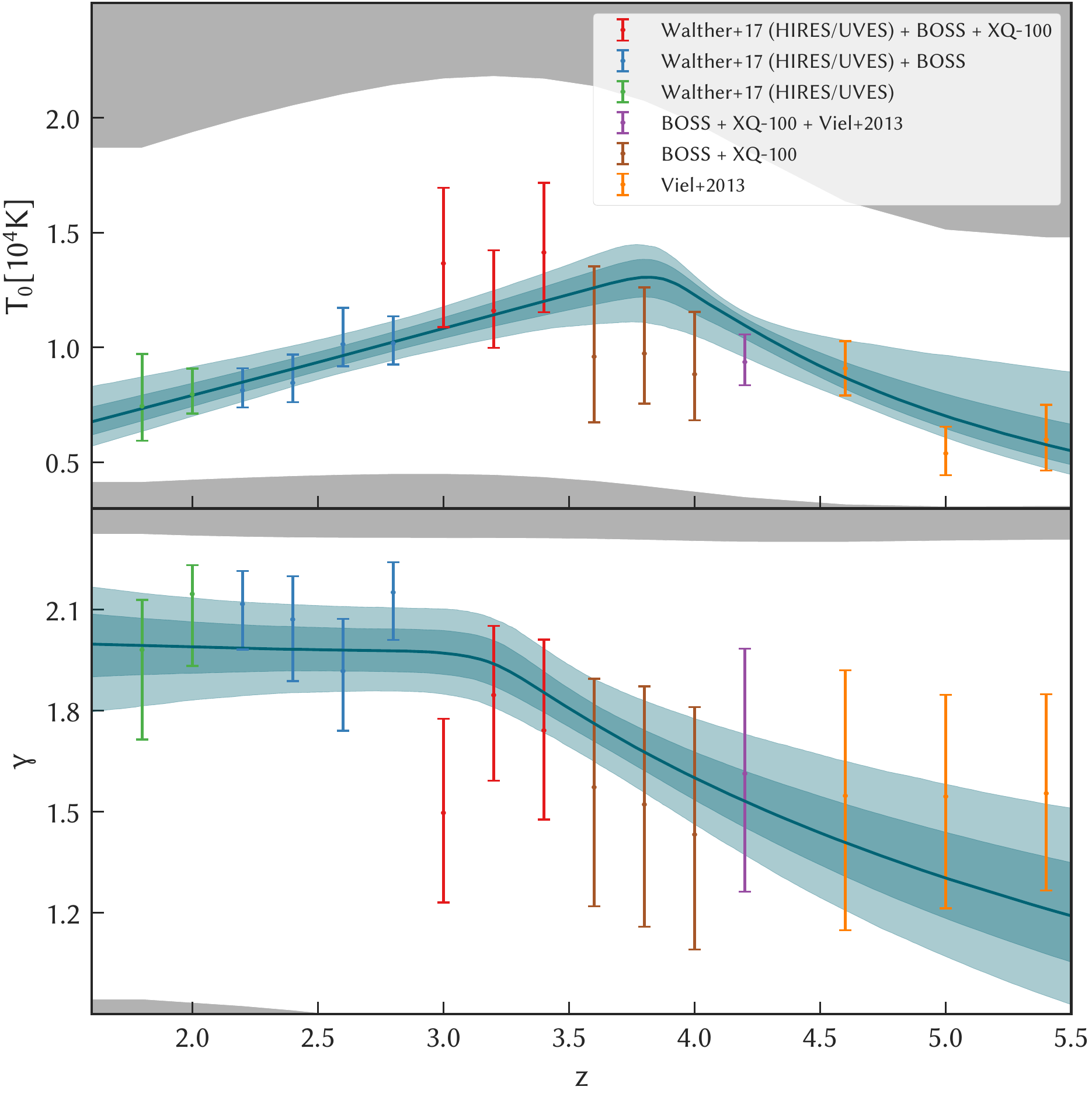}}}
  \endminipage
  \hfill
  \vskip -0.2cm
\caption{\small Observational Probes of Reionization: II.
\textit{Left:} Fluctuations in the \HI\ ionizing background for a galaxies+AGN source model. The map shows the \HI\ photoionization 
rate at $z=5.6$ in a slice of thickness 6.25 ${\rm Mpc}\slash h$ for a UV background model where AGNs produce 50\% of the global 
average \HI\ photoionization rate. The clustering of AGNs leads to large-scale variations in the mean free path of ionizing photons 
that amplify fluctuations in the ionizing background. (From \citep{daloisio17}.)
\textit{Center:} Ly$\alpha$ transmission spikes towards the $z=7.1$ quasar 
ULAS J1120+0641.
(From \citep{barnett17}.)
\textit{Right:} Evolution of the IGM temperature at mean density $T_0$ and $\rho-T$ relation
slope $\gamma$ as a function of redshift. (From \citep{walther19}.)
} 
\label{figEOR2}
\end{figure}

Figures \ref{figEOR1} and \ref{figEOR2} depict some of the available observational probes of the epoch of reionization. These include observations of the Gunn-Peterson absorption troughs indicating that the Universe was mostly reionized by $z\sim 6$ \citep{fan06,becker15}, 
the evolution of the cosmic SFR density at high redshift \citep{oesch17}, and the evolution of the IGM mean temperature at $z\lta 5.5$ 
\citep{walther19}. In addition, the sensitivity of 
\Lya\ emission to intervening neutral gas is also being used to explain the dearth of \Lya\ emitters (LAEs) observed 
at $z>6.5$ \citep{ouchi10,schenker14}. Recent {\it Planck} CMB anisotropy data together with  measurements of the 
kinetic Sunyaev-Zel'dovich effect by the Atacama Cosmology and the South Pole Telescopes constrain the redshift where the 
volume-averaged neutral fraction drops below 50\% to lie between $z=7.3$ and 10.5 ($2\sigma$), and yield an upper limit to the width 
of the reionization period of $\Delta z <2.8$ \citep{planck16}.

\subsection{The reionization equation}

Cosmic reionization involves a complex interplay between the abundance, clustering, spectrum, and LyC radiation leakage of
photoionizing sources, and the density, clumpiness, and temperature of intergalactic gas. Detailed simulations of reionization 
must be informed by insight from semi-empirical modeling and theoretical calculations that provide a proper accounting of the volume-averaged production and 
absorption of ionizing radiation in an expanding medium. 
Below, we give an example of how theory informs computation.

The general idea of a gradual reionization process driven by a steadily increasing UV photon production rate can be cast into a simple quantitative framework by integrating the ``reionization equation" \citep{madau17}
\begin{equation}
{d\langle x_\nHII\rangle\over dt} = {\langle \dot n_{\gamma}\rangle \over \langle n_\nH \rangle (1+\langle \kappa^{\rm LLS}_{\nu_L}\rangle/\langle \kappa^{\rm IGM}_{\nu_L}\rangle)}
-{\langle x_\nHII\rangle \over \bar t_{\rm rec}}.
\label{eq:dQdt}
\end{equation}
for the time evolution of the volume-averaged hydrogen ionized fraction
$x_\nHII$. Here, $\langle {\dot n}_{\gamma} \rangle$ is the emission rate {\it into the IGM} of ionizing photons per unit proper volume, $\langle n_\nH\rangle$ is the cosmological mean proper density of hydrogen, 
$\langle \kappa^{\rm LLS}\rangle^{-1}$ and $\langle \kappa^{\rm IGM}\rangle^{-1}$ are the photon proper mean free path at the hydrogen Lyman edge caused by the 
Lyman-limit systems (LLSs) and the diffuse IGM, respectively, and
$\bar t_{\rm rec}$ is an effective recombination timescale in a clumpy IGM.\footnote{Photon losses by LLSs are also due to radiative recombinations. In analytical models 
it is standard practice, however, to include the effect of LLSs as a reduction in the source term through the finite mean free path of ionizing radiation. 
Three different quantities -- the escape fraction, the gas clumping factor, and the mean free path -- are then used to describe what are essentially 
radiative recombinations in the ISM, the IGM, and the LLSs. Numerical simulations of cosmic reionization show that such a separation into distinct regimes 
may indeed be reasonable \citep{kaurov15}.}~ The above equation assumes a  photon mean free path that is much smaller than the horizon size (so that the mean specific intensity relaxes to the source function), and 
an early IGM that is organized in three main phases -- a uniform ionized medium, a uniform neutral medium, and the LLSs \citep{madau17}.

\begin{figure*}
\vskip -0.58cm 
\centering
\includegraphics[width=0.49\textwidth]{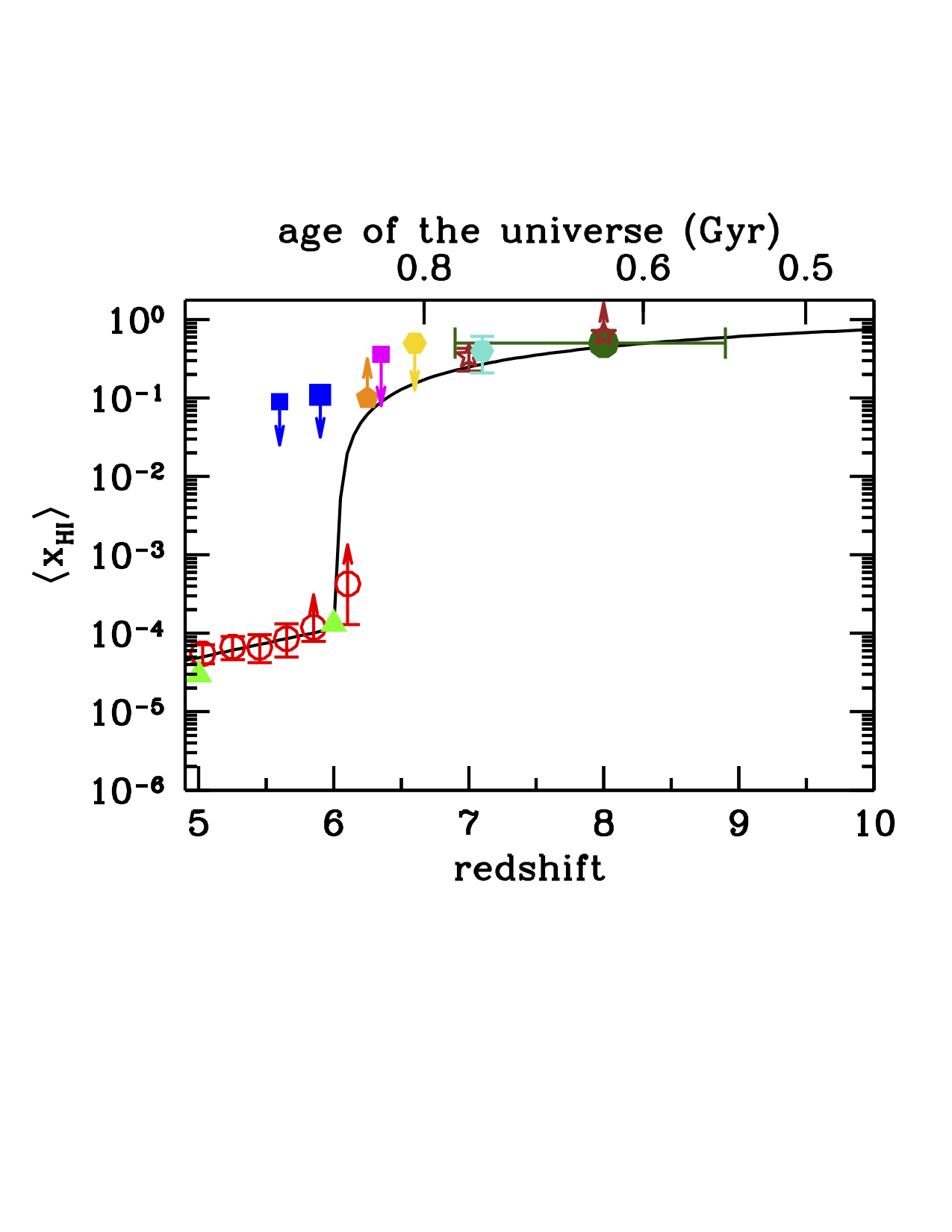}
\includegraphics[width=0.49\textwidth]{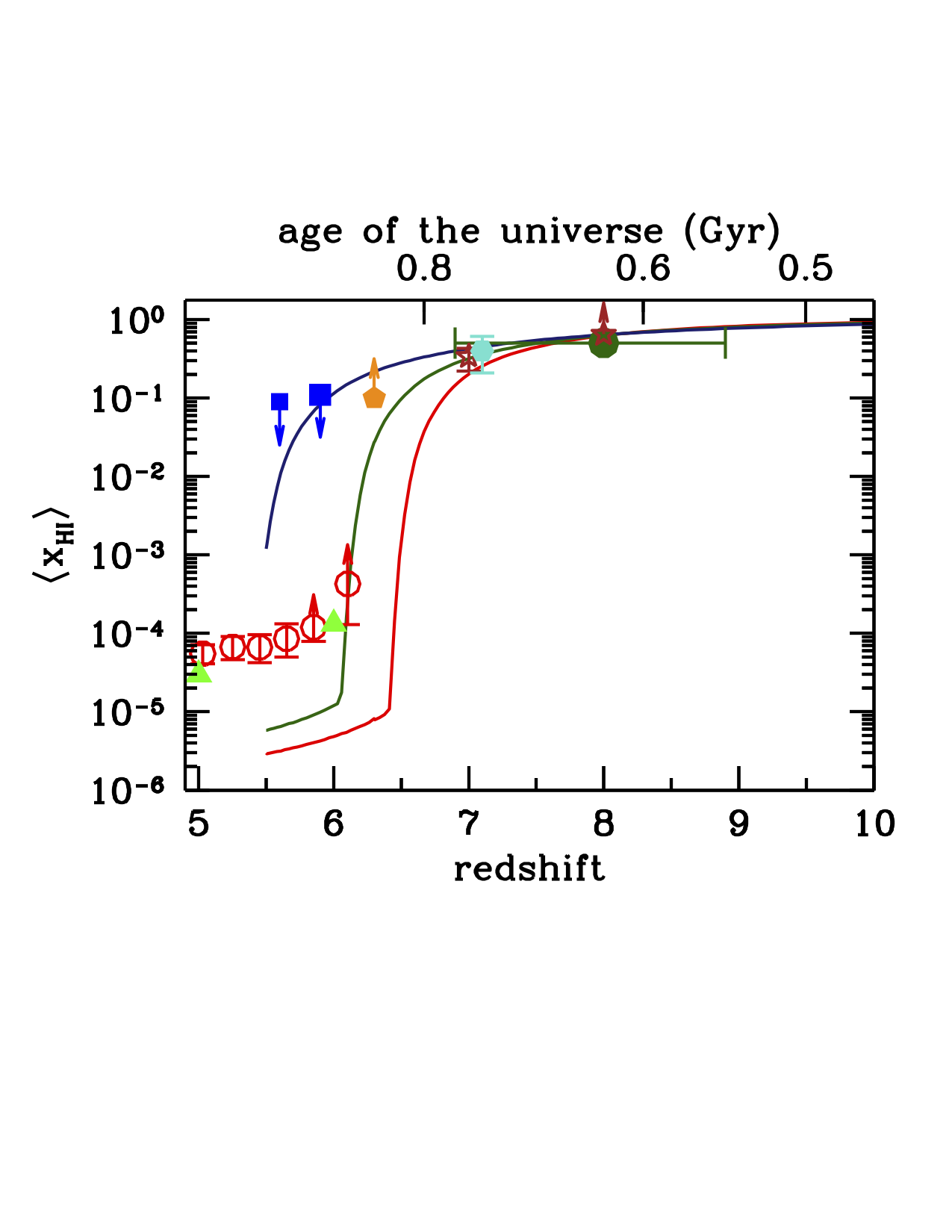}
\vskip -2.0cm
\caption{\small Average hydrogen neutral fraction, $\langle x_\nHI\rangle$, of a multi-phase IGM.
{\it Left panel}: Reionization histories predicted by integrating Equation (\ref{eq:dQdt}) with a constant emission rate of ionizing photons per hydrogen
atom, $\langle \dot n_{\gamma}\rangle/\langle n_\nH \rangle=3$ phot Gyr$^{-1}$.
The data points represent constraints on the ionization state of the IGM from: the Gunn-Peterson optical depth at $5<z<6.1$ (red open circles,
\citealt{fan06}), measurements of the \Lya\ forest opacity combined with hydrodynamical simulations (green triangles, \citealt{bolton07}),
the dark pixel statistics at $z=5.6$ and $z=5.9$ (blue squares, \citealt{mcgreer15}),
the gap/peak statistics at $z=6.32$ (magenta square, \citealt{gallerani08}),
the damping wing absorption profiles in the spectra of quasars at $z=6.3$ (orange pentagon, \citealt{schroeder13})
and $z=7.1$ (turquoise hexagon, \citealt{greig17}), the redshift-dependent prevalence of LAEs in narrow-band surveys at $z=7$ and $z=8$
(firebrick stars, \citealt{schenker14}) and their clustering properties at $z=6.6$ (gold hexagon, \citealt{ouchi10}).
The limit on the redshift at which $\langle x_\nHI\rangle=0.5$,
extracted from {\it Planck}'s CMB analysis \citep{planck16}, is plotted as the green dot.
{\it Right panel:} Reionization histories in three radiation-hydrodynamic simulations of the 
Reionization of Cosmic Hydrogen (SCORCH) project \citep{doussot17} that have different functional forms for the luminosity-weighted 
escape fraction $f_{\rm esc}$.  
}
\vskip -0.2cm
\label{figQ}
\end{figure*}

It is this ODE that statistically describes the transition from a neutral universe to a fully
ionized one, turning reionization into a {\it photon-counting exercise} in which the growth rate of \HII\ regions is equal to the rate at which ionizing photons are
produced minus the rate at which they are consumed by radiative recombinations in the IGM.
Before reionization, at $z>6$, UV photons carve out cosmological \HII\ regions in the uniform component that expand in an otherwise largely neutral phase. 
At $z < 6$, however, the diffuse IGM is observed to be highly ionized, with only a small residual amount of neutral hydrogen set by the balance between 
radiative recombinations and photoionizations from a nearly uniform UV radiation background, and provides negligibly small \HI\ photoelectric absorption. 
The continuum opacity is instead dominated by the optically thick (to ionizing radiation) LLSs, high density regions that trace non-linear and collapsed structures, 
occupy a small portion of the volume, and are able to keep a significant fraction of their hydrogen in neutral form 
\citep[e.g.][]{miralda00,gnedin06,furlanetto09,haardt12}. The integration of Equation (\ref{eq:dQdt}) therefore
provides a link between the pre-overlap and post-overlap phases of the reionization process.

Figure \ref{figQ} shows the reionization history obtained by numerically integrating Equation (\ref{eq:dQdt}) assuming a constant emission rate of ionizing
photons per hydrogen atom, $\langle \dot n_{\gamma}\rangle/\langle n_\nH \rangle=3$ phot Gyr$^{-1}$. 
The theoretical curve is consistent with a number of observational constraints on the ionization
state of the $z>5$ universe, from the redshift-dependent prevalence of LAEs in narrow band surveys at $z=$7--8 \citep{schenker14},
to the damping wing absorption profiles measured in the spectra of $z=6.3$ \citep{schroeder13} and $z=7.1$ quasars \citep{greig17}.
While the pre-overlap stages extend over a considerable range of redshift, the phase of overlap, indicated by the sudden drop in the neutral gas fraction at $z\simeq 6$,
is clearly defined even in the presence of the LLSs: $\langle x_\nHI\rangle$ decreases by more than 2 orders of magnitude over a fraction of the then Hubble time.
Such a dramatic transition marks the epoch when the photon mean free path becomes determined by the LLSs rather than by the typical size of \HII\ regions.
Note how current state-of-the-art radiation-hydrodynamic simulations of reionization produce a poor fit to the
observations of the Gunn-Peterson optical depth in the spectra of
Sloan Digital Sky Survey (SDSS) quasars at $5<z<61$ \citep{fan06}, as well as to estimates of
the ionized fraction at $z=5$ and $6$ based on a combination of hydrodynamical simulations with
measurements of the \Lya\ forest opacity \citep{bolton07}. The disagreement is an artifact of current reionization simulations underestimating the amount of UV absorption in the post-reionization Universe, as LLSs remain effectively unresolved.

\section{Concluding remarks}

We have reviewed our current understanding of early structure formation, from first stars to the first black holes emerging at high redshift, and
have addressed  the impact of early star forming galaxies and early accreting massive black holes on the reionization of the Universe. We have
highlighted the importance of radiative feedback in determining various key aspects of first objects, such as the masses of first stars and
the growth rate of the black holes arising from them. Radiative feedback, in particular dissociation of molecular hydrogen by the Lyman-Werner
radiation produced by early star forming galaxies, plays a key role also in models for massive black hole seed formation ($M_{BH} >10^4 M_{\odot}$ by direct collapse in metal-free atomic cooling halos. Likewise, thermodynamical  feedback from accretion and mergers could provide the required heating to avoid fragmentation in the most rapidly growing atomic cooling halos. Alternative models for direct collapse, such as from major mergers of the first metal-enriched massive galaxies at $z \sim 8-10$, actually rely exclusively on thermodynamical feedback from shocks and compressional heating
to stifle widespread fragmentation, and can give rise to much larger black hole seeds, of order a billion solar mass. In direct collapse models,
a paramount role is played by the physics of their likely precursors, accreting supermassive stars (SMS). We have thus covered in detail the latest developments in modeling SMSs, including rotation and the different regimes arising with accretion rates varying by orders of magnitude, according
to either the atomic cooling halo scenario and the galaxy merger-driven scenario. In doing so, we have reported on the possibility that collapse into
a massive black hole seed, at the highest accretion rates, occurs without prior formation of an SMS, the so-called {\it dark collapse} route.
The reionization of hydrogen during the first billion years of cosmic history is the last major phase transition of
baryonic matter in the Universe and involves the early production, propagation, and absorption of LyC radiation. The
full reionization of helium is believed to have occurred at a later time, when hard UV-emitting bright quasars and
other AGN became sufficiently abundant. While it is generally agreed that star-forming galaxies and the first
accreting seed black holes in their nuclei are the best candidates for providing the required ionizing power, there
are still many uncertainties on the relative contributions of these sources as a function of epoch. The resolution of
such uncertanties feels overdue and may soon be achieved by the large wavelength coverage, unique sensitivity, and
spectroscopic and imaging capabilities of the {\it JWST} together with ongoing and future experiments aimed at
measuring the redshifted 21-cm signal from neutral hydrogen during the epoch of first light.


\bibliographystyle{aps-nameyear3}

\bibliography{issi}

\end{document}